\documentclass[pra,twocolumn]{revtex4-2}
\usepackage{amsmath}
\usepackage{amsthm}
\usepackage{amssymb}
\usepackage{graphicx}
\usepackage{braket}
\usepackage{stmaryrd}
\usepackage{hyperref}
\usepackage{color}
\usepackage{dsfont}

\hypersetup{
    colorlinks,
    citecolor=blue,
    linkcolor=blue,
    urlcolor=blue
}

\newcommand{\dblbrace}[1]{\llbracket #1\rrbracket}

\def\xsum{\mathop{\sum\nolimits'}}

\theoremstyle{plain}
\newtheorem{lem}{Lemma}

\begin{document}
\title{A Unified Framework of Unitarily Residual Measures for Quantifying Dissipation}

\author{Tomohiro Nishiyama}
\email{htam0ybboh@gmail.com}
\affiliation{Independent Researcher, Tokyo 206-0003, Japan}

\author{Yoshihiko Hasegawa}
\email{hasegawa@biom.t.u-tokyo.ac.jp}
\affiliation{Department of Information and Communication Engineering, Graduate
School of Information Science and Technology, The University of Tokyo,
Tokyo 113-8656, Japan}
\date{\today}

\begin{abstract}

Open quantum systems are governed by both unitary and non-unitary dynamics, with dissipation arising from the latter. Traditional quantum divergence measures, such as quantum relative entropy, fail to account for the non-unitary oriented dissipation as the divergence is positive even between unitarily connected states. We introduce a framework for quantifying the dissipation by isolating the non-unitary components of quantum dynamics. We define equivalence relations among hermitian operators through unitary transformations and characterize the resulting quotient set. By establishing an isomorphism between this quotient set and a set of real vectors with ordered components, we induce divergence measures that are invariant under unitary evolution, which we refer to as the unitarily residual measures. These unitarily residual measures inherit properties such as monotonicity and convexity and, in certain cases, correspond to classical information divergences between sorted eigenvalue distributions. Our results provide a powerful tool for quantifying dissipation in open quantum systems, advancing the understanding of quantum thermodynamics.

\end{abstract}

\maketitle

\section{Introduction}

The dynamics of isolated quantum systems are governed by unitary transformations, which are reversible and do not result in thermodynamic dissipation. In contrast, open quantum systems cannot be fully described by unitary transformations alone \cite{Breuer:2002:OpenQuantum}. Naturally, the dissipation in open quantum systems is expected to arise from the non-unitary components, since unitary operations do not contribute to dissipation. In stochastic thermodynamics \cite{Seifert:2012:FTReview,VandenBroeck:2015:Review}, divergences play a central role in quantifying dissipation. Specifically, the Kullback-Leibler divergence quantifies entropy production by comparing the probabilities of forward and backward trajectories \cite{Parrondo:2009:Entropy,Roldan:2012:EntropyProduction,Seifert:2012:FTReview}. Similarly, in quantum thermodynamics, quantum divergences, such as quantum relative entropy, are important to quantify dissipation \cite{Esposito:2010:EntProd,Reed:2014:Landauer,Landi:2021:EPReview}. 
However, conventional quantum divergence measures remain positive even under purely unitary transformations; in other words, the divergence between states that can be transitioned to via a unitary operator does not vanish. Therefore, we need divergence measures that are independent of unitary components; the divergence measure between density operators connected by a unitary transformation should be zero.
Regarding the quantum Markov process described by the Lindblad equation, Refs.~\cite{Vu:2021:GeomBound} and \cite{PhysRevLett.127.190601} introduced the total variation distance and the Kullback-Leibler divergence between sorted eigenvalues of density operators and derived speed limits for the entropy production. Reference~\cite{Vu:2021:GeomBound} showed that the Kullback-Leibler divergence between sorted eigenvalues is equal to the minimum of the quantum relative entropy between a unitary transformed density operator and another density operator. Reference~\cite{van2023thermodynamic} showed that the same relation holds between the total variation distance and the trace distance.

\begin{figure}
    \centering
    \includegraphics[width=1.0\linewidth]{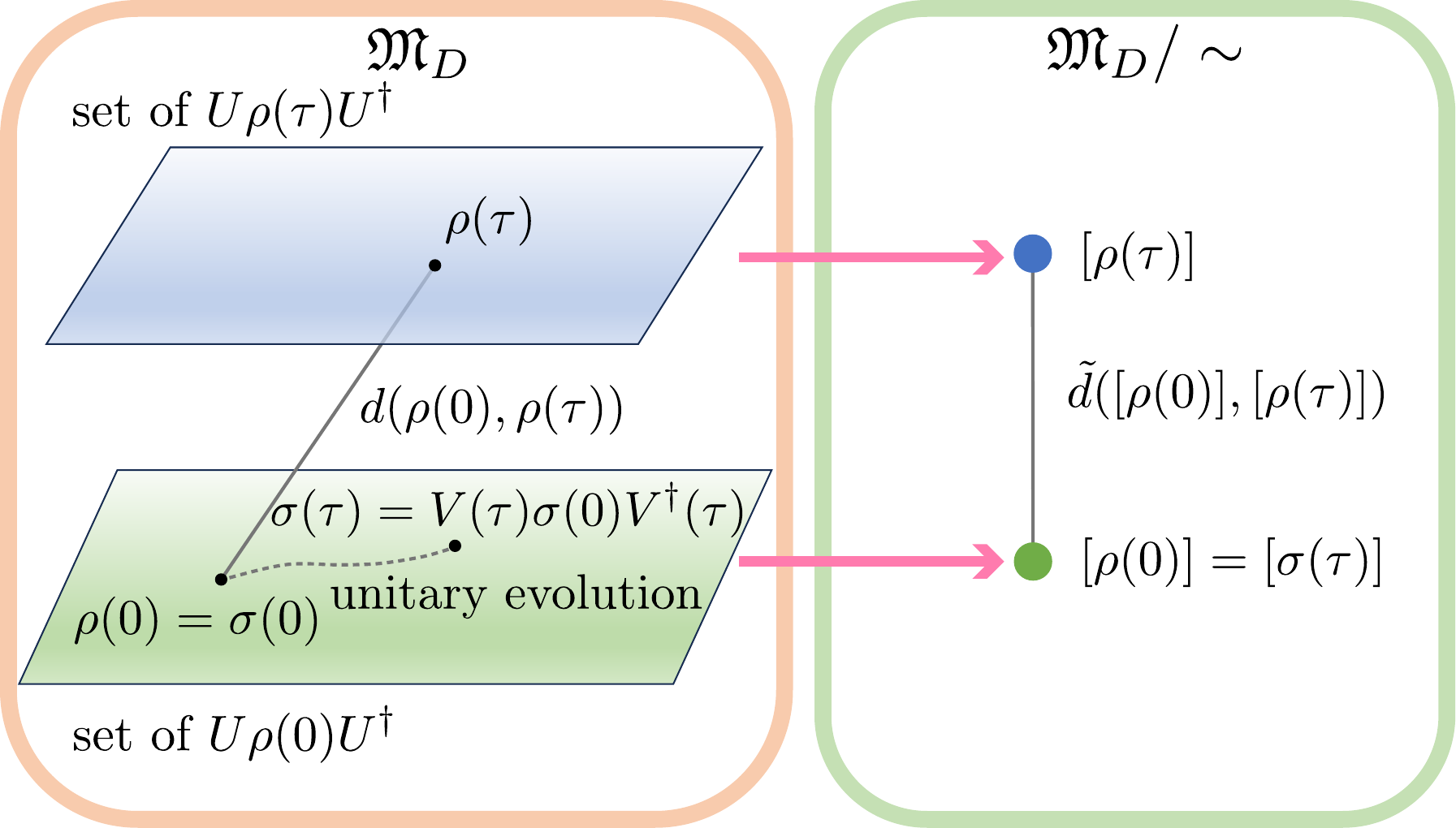}
    \caption{Illustration of equivalence classes, quotient set $\mathfrak{M}_D/\sim$ and unitarily residual measure $\widetilde{d}$. Time evolution of $\rho(t)$ comprises dissipation and $\sigma(t)$ unitarily evolves by $V(t)$. $U$ is an arbitrary unitary operator.
    Since $[\sigma(t)]$ stays single point in the quotient set, the unitarily residual measure quantifies the effect of dissipation. 
    }
    \label{fig:concept}
\end{figure}

The main aim of  this study  is to provide a unified framework for quantifying the effect of dissipation by non-unitary components.
We define equivalence relations between hermitian operators via unitary transformations and its quotient set. 
By identifying all quantum states that can be transitioned to via unitary transformations as a single point (Fig.~\ref{fig:concept}), we can isolate the effects of non-unitary (dissipative) time evolution in the quotient set. 
We show that isomorphism exists between the quotient set and a set of real vectors whose components are arranged in non-descending order. We also show that divergence measures are naturally induced in the quotient set from the quantum divergences between density operators (Fig.~\ref{fig:concept}), which we refer to as \textit{unitarily residual measures}. Under the assumption [cf. Eq.~\eqref{eq:assumption2}],
we show that the unitarily residual measure inherits fundamental properties such as monotonicity and convexity of the original quantum divergence. Regarding the monotonicity, when the original quantum divergence is monotonic with respect to completely-positive trace-preserving (CPTP) map, the unitarily residual measure is monotonic with stochastic map of eigenvalues.
For certain examples, the unitarily residual measures can be written as the classical information divergence between probability distributions of eigenvalues arranged in non-descending order (Table~\ref{tab:summary_metric}). These results allow us to write quantum speed limits on dissipation in semi-classical form. 
As an application, we show speed limits on unitarily residual measures in non-hermitian dynamics and we show speed limits for the purity.
\begin{table*}
    \centering
\begin{tabular}{|c|c|}
\hline 
Quantum divergence & Unitarily residual measure\tabularnewline
\hline 
\hline 
\begin{tabular}{c}
    Bures angle \\
    $\displaystyle \mathcal{L}_D(\rho, \sigma)=\mathrm{arccos}\left(\sqrt{\mathrm{Fid}(\rho,\sigma)}\right)$
\end{tabular}
&
\begin{tabular}{c}
     Bhattacharyya (arccos) distance \\
     $\displaystyle \widetilde{\mathcal{L}}_D([\rho], [\sigma])=\mathrm{arccos}\left(\xsum_{i} \sqrt{p_i q_i}\right)$
\end{tabular}
\tabularnewline
\hline 
\begin{tabular}{c}
     Trace distance \\
     $\displaystyle \mathcal{T}(\rho, \sigma)= \frac{1}{2} \|\rho-\sigma\|_1$ 
\end{tabular}
& 
\begin{tabular}{c}
    Total variation distance \\
    $\displaystyle \widetilde{\mathcal{T}}([\rho], [\sigma])=\frac{1}{2}\xsum_{i} |p_i - q_i|$
\end{tabular}
\tabularnewline
\hline 
\begin{tabular}{c}
    Petz-R\'{e}nyi relative entropy \\
    $\displaystyle D_\alpha(\rho\|\sigma)=\frac{1}{\alpha-1}\ln\left(\mathrm{Tr}\left[\rho^\alpha \sigma^{1-\alpha}\right]\right)$
\end{tabular}
&
\begin{tabular}{c}
    R\'{e}nyi divergence \\
    $\displaystyle \widetilde{D}_\alpha([\rho]\,\|\,[\sigma])=\frac{1}{\alpha-1}\ln\left(\xsum_{i} p_i^\alpha q_i^{1-\alpha}\right)$
\end{tabular}
\tabularnewline
\hline
\begin{tabular}{c}
    Sandwiched R\'{e}nyi relative entropy\\
    $\displaystyle D^\prime_\alpha(\rho\|\sigma)=\frac{1}{\alpha-1}\ln\left(\mathrm{Tr}\left[\left(\sigma^{\frac{1-\alpha}{2\alpha}}\rho \sigma^{\frac{1-\alpha}{2\alpha}}\right)^\alpha\right]\right)$
\end{tabular}
&
\begin{tabular}{c}
    R\'{e}nyi divergence\\
    $\displaystyle \widetilde{D^\prime}_\alpha([\rho]\,\|\,[\sigma])=\frac{1}{\alpha-1}\ln\left(\xsum_{i} p_i^\alpha q_i^{1-\alpha}\right)$
\end{tabular}
\tabularnewline
\hline
\end{tabular}
    \caption{Examples of quantum divergences on a set of density operators $\mathfrak{M}_D$ and corresponding unitarily residual measures on the quotient set $\mathfrak{M}_D/\sim$. $\xsum$ denotes the sum of non-decreasing sequences $\{p_i\}$ and $\{q_i\}$, which are eigenvalues of density operators $\rho$ and $\sigma$, respectively [cf. Eq.~\eqref{eq:ascending_notation}].
}
    \label{tab:summary_metric}
\end{table*}

\section{Equivalence classes \label{sec:equivalent_classes}}
Let $\mathcal{H}$ be a Hilbert space with dimension $n$, and let $\mathfrak{L}(\mathcal{H})$ be a set of linear operators on $\mathcal{H}$. 
Let $\mathfrak{M}\subset \mathfrak{L}(\mathcal{H})$
be a set of hermitian operators. Since the dimension of hermitian operators are $n^2$, we write $\mathfrak{M}_{n^2}$ when we emphasize the dimension. We define an equivalence relation $\sim$ between hermitian operators $A, B\in \mathfrak{M}$
via unitary transformations:
\begin{align}
    A\sim B \; \text{if} \  ^{\exists} U \ \text{such that} \  U^\dagger U=\mathbb{I}, \; B = U A U^\dagger.
    \label{eq:equivalence_relation}
\end{align}
Equation~\eqref{eq:equivalence_relation} shows that any two states connected by a unitary transformation are considered equivalent.
For all $A,B,C\in\mathfrak{M}$, this relation satisfies the following three properties:
\begin{align}
    &A\sim A, \label{eq:A_sim_A}\\
    &A\sim B \ \text{if and only if} \ B\sim A,\label{eq:A_sim_B_sim_A}\\
    &\text{If} \ A\sim B \ \text{and} \ B\sim C \ \text{then} \ A\sim C.\label{eq:A_sim_B_sim_C}
\end{align}
One can easily verify these relations from the definition in Eq.~\eqref{eq:equivalence_relation}.
The equivalence relation naturally splits $\mathfrak{M}$ into equivalence classes:
\begin{align}
    [A]:= \{B\in \mathfrak{M}:\ B\sim A\}.
    \label{eq:equivalence_class}
\end{align}
From this definition, $[A]=[B]$ holds when $A\sim B$.
A set of equivalence classes is called the quotient set, which is denoted by $\mathfrak{M}/\sim$. 

We next show that the isomorphism exists between a quotient set and a set of real vectors whose components are arranged in non-descending order. For the sake of simplicity, we introduce the notation before the discussion. 
Let $\mathbf{x}^\uparrow$ be a sorted vector which is obtained by arranging the components of $\mathbf{x}\in\mathbb{R}^n$ in non-descending order (i.e., $x_1^\uparrow\le x_2^\uparrow\le \cdots \le x_n^\uparrow$).
Let ${\mathbb{R}^n}^\uparrow:=\{\mathbf{x}^\uparrow:\; \mathbf{x}\in\mathbb{R}^n \}$ be a set of sorted vectors. For $\mathbf{a}^\uparrow, \mathbf{b}^\uparrow\in {\mathbb{R}^n}^\uparrow$, we use the notation $\xsum$ to define the sum of vectors in non-descending order: 
\begin{align}
    \xsum_{i=1}^{n} F(a_i, b_i):=\sum_{i=1}^{n} F(a^\uparrow_i, b^\uparrow_i),
    \label{eq:ascending_notation}
\end{align}
where $F$ is an arbitrary function. 
In a similar way, we write the sum $\xsum_i a_i \ket{b_i}\bra{b_i}$ and $\xsum_i \ket{b_i}\bra{a_i}$ for the non-descending sequences $ a_1\le a_2\le \cdots \le a_n$ and $b_1\le b_2\le \cdots \le b_n$, where $\ket{a_i}$ denotes an eigenvector corresponds to the the $i$-th eigenvalue $a_i$ ($\ket{b_i}$ is defined analogously). Let $A=\sum_i a_i\ket{a_i}\bra{a_i}$ and $B=\sum_i b_i\ket{b_i}\bra{b_i}$ be spectral decompositions. Since the sorted vector space ${\mathbb{R}^n}^\uparrow$ is closed under addition $\mathbf{a}^\uparrow + \mathbf{b}^\uparrow$ and scalar multiplication $ k\mathbf{a}^\uparrow$ for a non-negative real number $k$, we similarly define addition and scalar multiplication for equivalent classes:
\begin{align}
    &[A]+[B]:=\left[\xsum_{i} (a_i+b_i)\ket{a_i}\bra{a_i}\right], \label{eq:def_equivalent_sum}\\
    &k[A]:=\left[k\xsum_{i} a_i\ket{a_i}\bra{a_i}\right], \; \text{for} \; k\geq 0. 
    \label{eq:def_equivalent_multiplication}
\end{align}
Note that the right-hand side of Eqs.~\eqref{eq:def_equivalent_sum} and~\eqref{eq:def_equivalent_multiplication} do not depend on the choice of orthonormal basis since $\sum_i\ket{a_i}\bra{v_i}$ is a unitary operator for arbitrary orthonormal basis $\{\ket{v_i}\}$. 
 We prove that relations $[A]=[B]$ and $\mathbf{a}^\uparrow=\mathbf{b}^\uparrow$ are equivalent. Since unitary operators preserve eigenvalues, $\mathbf{a}^\uparrow=\mathbf{b}^\uparrow$ holds if $[A]=[B]$. Conversely, if $\mathbf{a}^\uparrow=\mathbf{b}^\uparrow$, there exists a unitary operator $W_{BA} := \xsum_{i} \ket{b_i}\bra{a_i}$ such that $B = W_{BA} A W_{BA}^\dagger$. Hence, the equivalence follows. This equivalence allows us to identify these two mathematical sets as essentially the same.
The relationship between these mathematical structures can be expressed as $\mathfrak{M}_{n^2}/\sim \, \cong {\mathbb{R}^n}^\uparrow$.
This notation indicates that $\mathfrak{M}_{n^2}/\sim$ and ${\mathbb{R}^n}^\uparrow$ are isomorphic, meaning that the two sets have the same properties. 
The detailed proof of the isomorphism is shown in Appendix~\ref{sec:equivalence_eigen}.
By identifying the set of equivalence classes with ${\mathbb{R}^n}^\uparrow$, it becomes easier to understand intuitively the structure of $\mathfrak{M}_{n^2}/\sim$.

\section{Unitarily residual measures}
Let $\mathfrak{M}_D\subset \mathfrak{M}$ be a set of density operators.
Consider a real-valued function $d$ that satisfies the following axiom.
For all $\rho, \sigma \in\mathfrak{M}_D$, 
\begin{align}
    &d(\rho, \sigma)\geq 0, \;d(\rho, \sigma)=0 \; \text{if and only if} \; \rho=\sigma. \label{eq:axiom1}    
\end{align}
The function $d$ is a {\em metric} when $d$ satisfies additional two axioms. For all $\rho, \sigma, \chi \in\mathfrak{M}_D$, 
\begin{align}
    &d(\rho, \sigma)=d(\sigma, \rho), \label{eq:axiom2}\\
    &d(\rho, \sigma)\le d(\rho,\chi)+d(\chi, \sigma).
    \label{eq:axiom3}
\end{align}
Equation~\eqref{eq:axiom3} is the triangle inequality.
We impose unitary invariance on $d$. That is, 
for an arbitrary unitary operator $U$, 
\begin{align}
    d(U\rho U^\dagger, U\sigma U^\dagger)=d(\rho,\sigma).
    \label{eq:axiom_unitary}
\end{align}
For instance, the condition is satisfied by trace distance, Bures angle and quantum relative entropy.
We define a  \textit{unitarily residual measure}  $\widetilde{d}$ on the quotient set $\mathfrak{M}_D/\sim$ as:
\begin{align}
    &\widetilde{d}([\rho],[\sigma]):=\min_{U^{\dagger}U=V^{\dagger}V=\mathbb{I}}d(U\rho U^{\dagger},V\sigma V^{\dagger})\nonumber\\&=\min_{U^{\dagger}U=\mathbb{I}}d(U\rho U^{\dagger},\sigma)=\min_{U^{\dagger}U=\mathbb{I}}d(\rho,U\sigma U^{\dagger}),
    \label{eq:def_metric_quotient}
\end{align}
where the minimum is over all possible unitaries $U$ and $V$, and we use Eq.~\eqref{eq:axiom_unitary} in the last two equalities. From Eq.~\eqref{eq:equivalence_class}, the unitarily residual measure satisfies Eq.~\eqref{eq:axiom1} . 
If $d$ is a metric, then the unitarily residual measure $\widetilde{d}$ on the quotient set also forms a metric, which we refer to as the \textit{unitarily residual metric}.  
Equation~\eqref{eq:axiom2} follows from the symmetry of the definition Eq.~\eqref{eq:def_metric_quotient}, and Eq.~\eqref{eq:axiom3} follows from
\begin{align}
    &\widetilde{d}([\rho], [\sigma])\le \min_{U^\dagger U=\mathbb{I}}d(U \rho U^\dagger, \chi)+ \min_{V^\dagger V
    =\mathbb{I}}d(\chi, V \sigma V^\dagger)\nonumber\\
    &=\widetilde{d}([\rho],[\chi])+\widetilde{d}([\chi],[\sigma]).
    \label{eq:triangle_property}
\end{align}
From the definition Eq.~\eqref{eq:def_metric_quotient}, the unitarily residual measure satisfies the following property:
\begin{align}
    \widetilde{d}([\rho], [\sigma])\le d(\rho,\sigma).
    \label{eq:quotient_ineq}
\end{align}

\subsection{Monotonicity and convexity}
Monotonicity and convexity are fundamental properties of quantum divergences. 
For a CPTP map $\mathcal{E}(\bullet)$ the monotonicity is defined as
\begin{align}
    d(\rho, \sigma)&\geq d(\mathcal{E}(\rho), \mathcal{E}(\sigma)).
    \label{eq:monotonicity_distance}
\end{align}
The condition is satisfied by trace distance, Bures angle and quantum relative entropy.
For non-negative real numbers $\{\lambda_i\}$ such that $\sum_i \lambda_i=1$, the convexity is defined as
\begin{align}
    \sum_i \lambda_i d(\rho_i, \sigma)\geq  d\left(\sum_i \lambda_i \rho_i, \sigma\right).
    \label{eq:convexity_distance}
\end{align}
The condition is satisfied by trace distance and quantum relative entropy. 
Letting $\rho=\sum_i p_i\ket{p_i}\bra{p_i}$ and $\sigma=\sum_j q_j\ket{q_j}\bra{q_j}$,  
we prove that the unitarily residual measures inherit these properties under the
additional assumption:
\begin{align}
    \widetilde{d}([\rho],[\sigma])&=d\left(\xsum_{i} p_i \ket{p_i}\bra{p_i}, \xsum_{i} q_i \ket{p_i}\bra{p_i}\right).
     \label{eq:assumption2}
\end{align}
Note that the right-hand side of Eq.~\eqref{eq:assumption2} does not depend on the choice of orthonormal basis from unitary invariance of $d$ [Eq.~\eqref{eq:axiom_unitary}]. This condition implies that $d(\rho,\sigma)$ is minimized when two density operators can be diagonalized by the same eigenvectors with eigenvalues $\{p_i^\uparrow\}$ and $\{q_i^\uparrow\}$. In the next section, we will verify the condition Eq.~\eqref{eq:assumption2} for all the quantum divergences in Table~\ref{tab:summary_metric}, which are widely used.

Since $\mathfrak{M}_{n^2}/\sim \, \cong {\mathbb{R}^n}^\uparrow$, we write $\mathbf{p}^\uparrow$ instead of $[\xsum_i p_i \ket{p_i}\bra{p_i}]$ for the sake of simplicity. 
For $[\rho]={\mathbf{p}}^\uparrow$, we define a stochastic map of eigenvalues $ \widetilde{\mathcal{E}}:\mathfrak{M}_D/\sim \,\rightarrow \mathfrak{M}_D^\prime/\sim$ as
\begin{align}
    \widetilde{\mathcal{E}}({\mathbf{p}}^\uparrow):={\left(T\mathbf{p}^\uparrow\right)}^\uparrow,
    \label{eq:def_stochastic_map}
\end{align}
where $T$ is a stochastic matrix (i.e, $\sum_i T_{ij}=1$ for all $i$ and $\{T_{ij}\}$ are all non-negative). 
If $d$ satisfies monotonicity Eq.~\eqref{eq:monotonicity_distance}, the unitarily residual measure is monotonically decreasing under $\widetilde{\mathcal{E}}$:
\begin{align}
    \widetilde{d}([\rho], [\sigma])\geq \widetilde{d}(\widetilde{\mathcal{E}}([\rho]), \widetilde{\mathcal{E}}([\sigma])).
    \label{eq:monotonicity_induced_distance}
\end{align}
If $d$ satisfies convexity Eq.~\eqref{eq:convexity_distance}, $\widetilde{d}$ also satisfies convexity: 
\begin{align}
    &\sum_i \lambda_i \widetilde{d}([\rho_i],[\sigma])\geq \widetilde{d}\left(\sum_i \lambda_i [\rho_i], [\sigma]\right),
    \label{eq:convexity_induced_distance}
\end{align}
where operations of equivalent classes are defined in Eqs.~\eqref{eq:def_equivalent_sum} and~\eqref{eq:def_equivalent_multiplication}.
The proofs of monotonicity and convexity are shown in  Appendix~\ref{sec:proof_monotonicity} and~\ref{sec:proof_convexity}.
One can prove the convexity with respect to $\sigma$ and the joint convexity
$\lambda d(\rho_1,\sigma_1) +  (1-\lambda) d(\rho_2,\sigma_2)\geq d(\lambda \rho_1+(1-\lambda) \rho_2, \lambda \sigma_1+(1-\lambda) \sigma_2)$
are also inherited. These inheritances are the  first  main results of  this study .

\subsection{Examples of unitarily residual measures}
Due to the isomorphism between $[\rho]$ and $\mathbf{p}^{\uparrow}$, the unitarily residual measure is expected to correspond to the classical information divergence between the probability distributions $\mathbf{p}^{\uparrow}$ and $\mathbf{q}^{\uparrow}$. We demonstrate that this correspondence holds in several important examples shown below.

One example of the metric $d$ is the Bures angle, which is widely employed in the literature~\cite{Nielsen:2011:QuantumInfoBook}:
\begin{align}
    \mathcal{L}_D(\rho, \sigma):= \arccos\left[\sqrt{\mathrm{Fid}(\rho, \sigma)}\right],
    \label{eq:L_D_def}
\end{align}
where $\mathrm{Fid}(\rho, \sigma)$ is the quantum fidelity: 
\begin{align}
    \mathrm{Fid}(\rho,\sigma):=\left(\mathrm{Tr}\left[\sqrt{\sqrt{\rho}\sigma\sqrt{\rho}}\right]\right)^{2}.
    \label{eq:fidelity_def}
\end{align}
The unitarily residual metric is the Bhattacharyya (arccos) distance~\cite{Bhattacharyya:1946:Divergence} between $\mathbf{p}^\uparrow$ and $\mathbf{q}^\uparrow$: 
\begin{align}
    \widetilde{\mathcal{L}}_D([\rho], [\sigma])=\mathrm{arccos}\left(\xsum_{i} \sqrt{p_i q_i}\right).
    \label{eq:dsim_bures}
\end{align}
The derivation of this relation is shown in  Appendix~\ref{sec:deriv_dsim_bures}.  
Another example of the metric $d$ is the trace distance, which is defined by 
\begin{align}
    \mathcal{T}(\rho, \sigma):= \frac{1}{2} \|\rho-\sigma\|_1,
    \label{eq:def_trace_dist}
\end{align}
where $ \|X\|_1:=\mathrm{Tr}[\sqrt{X^\dagger X}]$.
The unitarily residual metric of the trace distance is given by the total variation distance between $\mathbf{p}^\uparrow$ and $\mathbf{q}^\uparrow$:
\begin{align}
    \widetilde{\mathcal{T}}([\rho], [\sigma])=\frac{1}{2}\xsum_{i} |p_i - q_i|.
    \label{eq:ineq_eigen_trace}
\end{align}
Equation.~\eqref{eq:ineq_eigen_trace} was shown in Ref.~\cite{van2023thermodynamic}  (see  Appendix~\ref{sec:deriv_trace_eigen}).

As the last example, we consider 
two kinds of R\'{e}nyi relative entropy for $\alpha\in(0,1) \cup (1,\infty)$.
The first is 
the Petz-R\'{e}nyi relative entropy~\cite{Petz:1986:QuasiEntropies}:
\begin{align}
    D_\alpha(\rho||\sigma):=\frac{1}{\alpha-1}\ln\left(\mathrm{Tr}\left[\rho^\alpha \sigma^{1-\alpha}\right]\right).
    \label{eq:Quantum_renyi_def}
\end{align}
The second is the sandwiched R\'{e}nyi relative entropy~\cite{muller2013quantum, wilde2014strong}:
\begin{align}
    D^\prime_\alpha(\rho||\sigma):=\frac{1}{\alpha-1}\ln\left(\mathrm{Tr}\left[\left(\sigma^{\frac{1-\alpha}{2\alpha}}\rho \sigma^{\frac{1-\alpha}{2\alpha}}\right)^\alpha\right]\right).
    \label{eq:sandwiched_renyi_def}
\end{align}
In the limit $\alpha\rightarrow 1$, the Petz-R\'{e}nyi relative entropy  and the sandwiched R\'{e}nyi relative entropy  reduce to the quantum relative entropy $D(\rho||\sigma):=\mathrm{Tr}\left[\rho\ln\rho-\rho\ln\sigma\right]$:
\begin{align}
    \lim_{\alpha\rightarrow 1}D_\alpha(\rho||\sigma)=\lim_{\alpha\rightarrow 1}D^\prime_\alpha(\rho||\sigma)=D(\rho||\sigma). 
    \label{eq:quantum_relative_entropy}
\end{align}
The sandwiched R\'{e}nyi relative entropy for $\alpha=1/2$ is related to the fidelity from Eq.~\eqref{eq:fidelity_def}:
\begin{align}
    D^\prime_{\frac{1}{2}}(\rho||\sigma)=-\ln (\mathrm{Fid}(\rho,\sigma)).
\end{align}
The Petz-R\'{e}nyi relative entropy and the sandwiched R\'{e}nyi relative entropy are not metric since they do not satisfy Eqs.~\eqref{eq:axiom2} and~\eqref{eq:axiom3}. 
The unitarily residual measure of both Petz-R\'{e}nyi relative entropy and sandwiched R\'{e}nyi relative entropy is the R\'{e}nyi divergence~\cite{renyi1961measures} between $\mathbf{p}^{\uparrow}$ and $\mathbf{q}^{\uparrow}$:
\begin{align}
    \widetilde{D}_\alpha([\rho]\,\|\,[\sigma])=\frac{1}{\alpha-1}\ln\left(\xsum_{i} p_i^\alpha q_i^{1-\alpha}\right).
    \label{eq:induced_renyi_relative_entropy}
\end{align}
\begin{align}
    \widetilde{D^\prime}_\alpha([\rho]\,\|\,[\sigma])=\widetilde{D}_\alpha([\rho]\,\|\,[\sigma]).
    \label{eq:induced_sandwiched}
\end{align}
The derivation of Eqs.~\eqref{eq:induced_renyi_relative_entropy} and~\eqref{eq:induced_sandwiched} are shown in Appendix~\ref{sec:deriv_induced_renyi_relative_entropy} and~\ref{sec:deriv_induced_sandwiched}, respectively. 
In  the  limit $\alpha\rightarrow 1$, we obtain
\begin{align}
    \widetilde{D}([\rho]\,\|\,[\sigma]):=\lim_{\alpha\rightarrow 1}\widetilde{D}_\alpha([\rho]\,\|\,[\sigma])&=\xsum_{i} p_i\ln \frac{p_i}{q_i},
    \label{eq:induced_relative_entropy}
\end{align}
where the right-hand side is the Kullback-Leibler divergence.
Equation.~\eqref{eq:induced_relative_entropy} was shown in Ref.~\cite{PhysRevLett.127.190601}. 
The correspondences between quantum divergences and unitarily residual measures are summarized in Table~\ref{tab:summary_metric}. 
The first, third and forth  rows  in Table~\ref{tab:summary_metric} are the second main results in this study.  We verify Eq.~\eqref{eq:assumption2} for all quantum divergences in Table~\ref{tab:summary_metric}. These quantum divergences reduce to the corresponding classical divergences when two density operators commute. In the right-hand side of Eq.~\eqref{eq:assumption2}, two density operators commute and their eigenvalues are $\mathbf{p}^\uparrow$ and $\mathbf{q}^\uparrow$ for the same eigenvectors. Since right-hand sides of Eqs.~\eqref{eq:dsim_bures}, ~\eqref{eq:ineq_eigen_trace}, ~\eqref{eq:induced_renyi_relative_entropy} and~\eqref{eq:induced_sandwiched} are classical divergences between $\mathbf{p}^\uparrow$ and $\mathbf{q}^\uparrow$, it follows that all the examples in Table~\ref{tab:summary_metric} satisfy Eq.~\eqref{eq:assumption2}.

\section{Non-hermitian dynamics\label{sec:non_hermitian}}
As an example, we consider the non-hermitian dynamics governed by the non-hermitian Hamiltonian $\mathcal{H}$. We demonstrate speed limits for the skew-hermitian (dissipative) component  of the Hamiltonian.
 In general, $\mathcal{H}$ can be decomposed into
\begin{align}
    \mathcal{H}(t) = H(t) - i\Gamma(t),
    \label{eq:nonHermitianHamiltonian_decompose}
\end{align}
where $H(t)$ and $\Gamma(t)$ are hermitian operators. The second term in Eq.~\eqref{eq:nonHermitianHamiltonian_decompose} is the  dissipative component . Consider a density operator $\rho(t)$, whose time evolution is governed by 
\begin{align}
     \dot{\rho}(t)=-i(\mathcal{H}(t)\rho(t)-\rho(t)\mathcal{H}^{\dagger}(t)).
     \label{eq:nonHermitian_Heisenberg_eq}
 \end{align}
Equation~\eqref{eq:nonHermitian_Heisenberg_eq} reduces to the von Neumann equation when $\mathcal{H}(t)$ is hermitian. 
Let $\widehat{\rho}(t)$ be a normalized density operator defined as
\begin{align}
    \widehat{\rho}(t):= \frac{\rho(t)}{\mathrm{Tr}[\rho(t)]}.
    \label{eq:normalized_density}
\end{align}
For the normalized density operator, Eq.~\eqref{eq:nonHermitian_Heisenberg_eq} is modified as 
\begin{align}
     &\dot{\widehat{\rho}}(t)=-i(\mathcal{H}(t)\widehat{\rho}(t)-\widehat{\rho}(t)\mathcal{H}^{\dagger}(t))+ 2\braket{\Gamma}(t) \widehat{\rho}(t),
\label{eq:nonHermitian_Heisenberg_eq_normalized}
\end{align}
where $\braket{X}(t):=\mathrm{Tr}[X(t)\widehat{\rho}(t)]$ denotes a mean of $X(t)$.
Let $\dblbrace{X}(t):=\sqrt{\braket{X^{2}}(t)-\braket{X}(t)^{2}}$ be a standard deviation of  an hermitian operator  $X(t)$, and let $V(t):= e^{-i\mathbb{T}\int_{0}^{t} H(t)dt}$, where $\mathbb{T}$ is the time ordered product and we adopt the convention of setting $\hbar =1$.
The upper bound for the Bures angle $\mathcal{L}_D$ is given by
\begin{align}
    \int_{0}^{\tau} \dblbrace{\Gamma}(t) dt\geq \mathcal{L}_D( \widehat{\rho}(0) , V(\tau)^\dagger\widehat{\rho}(\tau)V(\tau)),
    \label{eq:main_result_MT_like_non_Hermitian}
\end{align}
where we assume $\int_{0}^{\tau} \dblbrace{\Gamma}(t) dt \le \pi/2$.
The details of the derivation are shown in  Appendix~\ref{subseq:deriv_nonhermitian_bures}. 
Applying Eq.~\eqref{eq:quotient_ineq} for Eq.~\eqref{eq:main_result_MT_like_non_Hermitian}, we obtain a Mandelstam–Tamm-type speed limit: 
\begin{align}
    \tau\ge\tau_{\min}:=\frac{\widetilde{\mathcal{L}}_D([\widehat{\rho}(0)], [\widehat{\rho}(\tau)])}{\overline{\dblbrace{\Gamma}(t)}}.
    \label{eq:main_result_eigen_non_Hemirian}
\end{align}
Here, 
$\overline{\bullet}:=\frac{1}{\tau}\int_{0}^{\tau}\bullet dt$ 
is the time average of the quantity over the duration $[0,\tau]$.
This is the  third  main result in  this study.  
Note that Eq.~\eqref{eq:main_result_eigen_non_Hemirian} holds between any time interval $[t_1, t_2]$ since we can choose arbitrary time as $t=0$. 
As in Eq.~\eqref{eq:main_result_eigen_non_Hemirian}, eigenvalues of the normalized density operator at time $t=0$ and $t=\tau$ are nearly equal with small standard deviation of $\Gamma(t)$. 
One can estimate the minimum time required for the eigenvalues of the density operator $\widehat{\rho}$ to vary from $\mathbf{p}$ to $\mathbf{p}^\prime$.
Reference~\cite{PhysRevA.104.052620} introduced the Mandelstam–Tamm quantum speed limit for the generalized standard deviation $\sqrt{\dblbrace{H}(t)^2+\dblbrace{\Gamma}(t)^2-i\braket{[H,\Gamma]}(t)}$. Our result, as shown in Eq.~\eqref{eq:main_result_eigen_non_Hemirian}, 
provides an upper bound that relies solely on dissipative component of the Hamiltonian. In the short time limit $t_2\rightarrow t_1=t$, Eq.~\eqref{eq:main_result_eigen_non_Hemirian} reduces to 
\begin{align}
    \dblbrace{\Gamma}(t) \geq \frac{\sqrt{I(t)}}{2},
    \label{eq:eigen_non_Hermitian_short_time}
\end{align}
where $I(t):=\sum_i p_i(t) \left(d_t\ln p_i(t) \right)^2$ is the Fisher information for 
$\widehat{\rho}(t)=\sum_i p_i(t)\ket{p_i(t)}\bra{p_i(t)}$
, and $d_t:=d/dt$. 
This bound is attained when all $\{\ket{p_i(t)}\}$ are eigenvectors of $\Gamma(t)$. The details of derivation and proof of equality condition are shown in  Appendix~\ref{subseq:proof_nonhermitian_tightness}. 
Combining $\sqrt{I(t)}\geq |d_t p_i(t)|/\sqrt{p_i(t)} \geq 2|d_t (\sqrt{p_i(t)})|$ with Eq.~\eqref{eq:eigen_non_Hermitian_short_time} and integrating from time $t=0$ to $\tau$, we obtain
\begin{align}
    \int_0^\tau \dblbrace{\Gamma}(t) dt\geq \int_0^\tau \left|d_t \left(\sqrt{p_i(t)}\right)\right| dt\geq \left|\sqrt{p_i(\tau)}-\sqrt{p_i(0)}\right|.
\end{align}
This inequality gives the possible time-varying range of eigenvalues of the density operator.
Let $\mathcal{P}(t):= \mathrm{Tr}[\widehat{\rho}(t)^2]=\sum_i p_i^2$ be a purity. 
From Eq.~\eqref{eq:main_result_eigen_non_Hemirian}, a speed limit for the purity follows:
\begin{align}
   2\sin\left(\int_{0}^{\tau} \dblbrace{\Gamma}(t) dt\right)\geq |\mathcal{P}(\tau)-\mathcal{P}(0)|.
    \label{eq:purity_non_hermitian}
\end{align}
The details of the derivation are shown in  Appendix~\ref{subseq:deriv_nonhermitian_purity}. 
This inequality gives a non-trivial bound when $\int_{0}^{\tau} \dblbrace{\Gamma}(t) dt < \pi/6$.
Reference~\cite{DelCampo:2013:OpenQSL} introduced a quantum speed limit for relative purity that incorporates the adjoint Lindblad superoperator. Reference~\cite{Uzdin:2016:SpeedLimits} presented a quantum speed limit for purity, where the upper bound is determined by the norm of the Lindblad jump operators. 
Our results, as presented in Eq.~\eqref{eq:purity_non_hermitian}, offer an upper bound that depends exclusively on the dissipative component of the Hamiltonian.
For a specific example of the non-hermitian dynamics, we examine a quantum Markov process which is described by the Lindblad equation. Let $\mathcal{L}$ be the Lindblad superoperator. The Lindblad equation is given by
\begin{align}
    \dot{\rho}(t)=\mathcal{L}\rho(t)=-i\left[H,\rho(t)\right]+\sum_{m=1}^{N_{C}}\mathcal{D}\left[L_{m}\right]\rho(t),
    \label{eq:Lindblad_def}
\end{align}
Here, $H$ is the Hamiltonian, and $L_m$ is the $m$th jump operator, with $m$ ranging from $1$ to $N_C$ (the number of channels). The term 
$\mathcal{D}[L]\rho:= L\rho L^{\dagger}-\frac{1}{2}\{L^{\dagger}L,\rho\}$ 
is the dissipator, which describes the interaction between the system and its environment. Equation~\eqref{eq:Lindblad_def} can also be expressed as
\begin{align}
    \dot{\rho}(t)=-i\left(H_{\mathrm{eff}}\rho(t)-\rho(t)H_{\mathrm{eff}}^{\dagger}\right)+\sum_{m=1}^{N_{C}}L_{m}\rho(t)L_{m}^{\dagger},
    \label{eq:Lindblad2}
\end{align}
where $H_{\text{eff}} := H - \frac{i}{2} \sum_m L_m^{\dagger} L_m$
is the effective Hamiltonian, which is a non-hermitian operator.
Equation~\eqref{eq:Lindblad2} shows that the time evolution of the Lindblad equation comprises two contributions; the continuous evolution driven by the effective Hamiltonian (the first term in Eq.~\eqref{eq:Lindblad2}) and the discrete jump dynamics triggered by the jump operators (the second term in Eq.~\eqref{eq:Lindblad2}).
In classical Markov processes, the classical dynamical activity plays a central role in trade-off relations such as speed limits and thermodynamic uncertainty relations \cite{Garrahan:2017:TUR,Shiraishi:2018:SpeedLimit,Terlizzi:2019:KUR,Hasegawa:2020:QTURPRL,Hasegawa:2023:BulkBoundaryBoundNC}. 
The classical dynamical activity is defined as 
\begin{align}
    \mathcal{A}(\tau) := \int_{0}^{\tau}\sum_{m=1}^{N_C}\mathrm{Tr}[L_{m}\rho(t)L_{m}^{\dagger}]dt.
    \label{eq:classical_DA_def}
\end{align}
The classical dynamical activity quantifies the activity of systems by the average number of jump events during $[0,\tau]$. 
Let $\mathcal{B}(\tau)$ represent the quantum generalization of $\mathcal{A}(\tau)$ from Eq.~\eqref{eq:classical_DA_def}, known as the quantum dynamical activity \cite{Hasegawa:2020:QTURPRL,Nishiyama:2024:ExactQDAPRE} 
(see Appendix~\ref{sec:QDA_def} for the expression of $\mathcal{B}(\tau)$).
While the classical dynamical activity quantifies the level of system activity based on jump statistics, $\mathcal{B}(\tau)$ evaluates the activity of open quantum systems over the interval $[0,\tau]$. In quantum dynamics, state changes can occur without jumps due to coherent dynamics. Therefore, $\mathcal{B}(\tau)$ accounts for contributions from both jump and coherent dynamics.
Specifically, for $0\le(1/2)\int_{0}^{\tau}\sqrt{\mathcal{B}(t)}/t\,dt\le\pi/2$, the quantum speed limit holds~\cite{Hasegawa:2023:BulkBoundaryBoundNC}:
\begin{align}
    \tau\ge\tau_{\min}:=\frac{2\mathcal{L}_D(\rho(0), \rho(\tau))}{\frac{1}{\tau}\int_0^\tau \sqrt{\mathcal{B}(t)}/t\,dt}.
    \label{eq:QSL_Lindblad}
\end{align}
From the result in Ref.~\cite{Nishiyama:2024:ExactQDAPRE}, when $H=0$, the quantum dynamical activity reduces to $\mathcal{A}(t)$.
In the following, we assume $[L_m, H]=\omega_m L_m$, where $\omega_m$ is the result of subtracting the energy after the transition from the energy before the transition.  
Let $X_I(t):=V(t)^\dagger X(t) V(t)$ be an interaction picture of operator $X(t)$ for $V(t)= e^{-iHt}$.
From $L_m V(t)=e^{-i\omega_m t}V(t)L_m $, it follows that $L_{I,m}(t)=e^{-i\omega_m t}L_m$ and 
\begin{align}
    \dot{\rho_I}(t)=\sum_{m=1}^{N_{C}}\mathcal{D}\left[L_{I, m}(t)\right]\rho_I(t)=\sum_{m=1}^{N_{C}}\mathcal{D}\left[L_{m}\right]\rho_I(t),
    \label{eq:Lindblad_interaction}
\end{align}
from Eq.~\eqref{eq:Lindblad_def}. Applying Eq.~\eqref{eq:QSL_Lindblad} for $H=0$ and $\rho_I(t)$, and using 
$\int_{0}^{\tau}\sum_{m}\mathrm{Tr}[L_{m}\rho_I(t)L_{m}^{\dagger}]dt=\int_{0}^{\tau}\sum_{m}\mathrm{Tr}[L_{I,m}(t)\rho_I(t)L_{I,m}^{\dagger}(t)]dt=\mathcal{A}(\tau)$, we obtain 
\begin{align}
    \tau\ge&\tau_{\min}:=\frac{2\mathcal{L}_{D}(\rho(0),V^{\dagger}(\tau)\rho(\tau)V(\tau))}{\frac{1}{\tau}\int_{0}^{\tau}\sqrt{\mathcal{A}(t)}/t\,dt}\nonumber\\\geq&\frac{2\widetilde{\mathcal{L}}_{D}([\rho(0)],[\rho(\tau)])}{\frac{1}{\tau}\int_{0}^{\tau}\sqrt{\mathcal{A}(t)}/t\,dt},
    \label{eq:QSL_Lindblad_residual}
\end{align}
where we use Eq.~\eqref{eq:quotient_ineq}.
Equation~\eqref{eq:QSL_Lindblad_residual} is the fourth main result of this study. 
Equation~\eqref{eq:QSL_Lindblad_residual} shows that
the minimum time required for the time evolution is bounded from below by
the distance between $\rho(0)$ and $\rho(\tau)$, quantified by their eigenvalues, 
and the classical dynamical activity $\mathcal{A}(t)$.
Equation~\eqref{eq:QSL_Lindblad_residual} offers an advantage over Eq.~\eqref{eq:QSL_Lindblad} because the denominator on the right-hand side involves solely the dynamical activity $\mathcal{A}(t)$, excluding the quantum dynamical activity $\mathcal{B}(t)$.

\section{Conclusion}
This work provides a unified framework for quantifying the effect of dissipation. For this purpose, we introduced the equivalence classes of hermitian operators via unitary transformations and their quotient set. We showed that isomorphism exists between the quotient set and a set of real vectors whose components are in non-descending order, and we showed that unitarily residual measure on the quotient set are naturally induced from quantum divergences between density operators. Under an appropriate assumption, we showed that the unitarily residual measures inherit the monotonicity and convexity of the original quantum divergences. In some examples, the unitarily residual measures can be written as a classical information divergence between probability distributions of sorted eigenvalues of density operators. 
As an application,
we derived speed limits on the unitarily residual measure and purity for the skew-hermitian component of the Hamiltonian in non-hermitian dynamics. 
  In addition, 
 we derived speed limits on the unitarily residual measure for the skew-hermitian component of the Hamiltonian and classical dynamical activity in non-hermitian dynamics.
Further study of the properties of the quotient set and unitarily residual measures would be the focus of our future work.

\appendix

\begin{widetext}

\section{Proofs of Properties with respect to quotient set and unitarily residual measures}
\subsection{Isomorphism between $\mathfrak{M}_{n^2}/\sim$ and ${\mathbb{R}^n}^\uparrow$\label{sec:equivalence_eigen}}
Let $f:\mathfrak{M}_{n^2}/\sim\rightarrow{\mathbb{R}^n}^\uparrow$ be a map such that
\begin{align}
    f([A]):=\mathbf{a}^\uparrow.
    \label{eq:def_isomorphism}
\end{align}
This map is well-defined because eigenvalues are invariant under unitary transformations.
For all $A, B\in \mathfrak{M}_{n^2}$, we show that 
\begin{align}
    &f([A]+[B])=f([A])+f([B]), \label{eq:iso_addition}\\
    &f(k[A])=kf([A]), \; \text{for} \; k\geq 0,
    \label{eq:iso_multiplication}
\end{align} 
hold and the map $f$ is bijective.
Equations~\eqref{eq:iso_addition} and ~\eqref{eq:iso_multiplication} follow from the definitions of Eqs.~\eqref{eq:def_equivalent_sum},~\eqref{eq:def_equivalent_multiplication},~\eqref{eq:def_isomorphism} and ${(\mathbf{a}^\uparrow+\mathbf{b}^\uparrow)}^\uparrow=\mathbf{a}^\uparrow+\mathbf{b}^\uparrow$.
Next, we prove that $f$ is bijective by proving it is both surjective and injective. From Eq.~\eqref{eq:def_isomorphism}, it follows that $f$ is surjective because $f(\mathfrak{M}_{n^2}/\sim) = {\mathbb{R}^n}^\uparrow$. As shown in Section~\ref{sec:equivalent_classes}, $[A] = [B]$ follows if $\mathbf{a}^{\uparrow} = \mathbf{b}^{\uparrow}$. Therefore, $f$ is injective. Combining both results, we proved that $f$ is bijective. Except for the condition $k\geq 0$, the map $f$ is a linear isomorphism between two vector spaces.

\subsection{Proof of monotonicity [Eq.~\eqref{eq:monotonicity_induced_distance}]\label{sec:proof_monotonicity}}
From assumption Eq.~\eqref{eq:assumption2}, we obtain 
\begin{align}
    \widetilde{d}([\rho],[\sigma])=\widetilde{d}(\mathbf{p}^\uparrow,\mathbf{q}^\uparrow)=d\left(\xsum_{j} p_j \ket{p_j}\bra{p_j}, \xsum_{j} q_j \ket{p_j}\bra{p_j}\right).
    \label{eq:induced_commute}
\end{align}
Let $\mathbf{r}:=T\mathbf{p}^\uparrow$ and $\mathbf{s}:=T \mathbf{q}^\uparrow$. Let $K_i:=\sum_j \sqrt{T_{ij}} \ket{r_i}\bra{p^\uparrow_j}=\sum^\prime_j \sqrt{T_{ij}} \ket{r_i}\bra{p_j}$ for $\{\ket{p_j}\}\in \mathcal{H}$ and $\{\ket{r_i}\}\in \mathcal{H}^\prime$. 
We obtain $\sum_i K_i \left(\xsum_{j} p_j \ket{p_j}\bra{p_j}\right) K_i^\dagger=\sum_i r_i \ket{r_i}\bra{r_i}$ and $\sum_i K_i \left(\xsum_{j} q_j \ket{p_j}\bra{p_j}\right) K_i^\dagger=\sum_i s_i \ket{r_i}\bra{r_i}$.
Since $\sum_i K_i^\dagger K_i=\mathbb{I}$ from $\sum_i T_{ij}=1$, the map $\mathcal{E}_K(\bullet):=\sum_i K_i \bullet K_i^\dagger$ is a CPTP map. Here 
$\mathbb{I}$ is an identity operator in $\mathfrak{L}(\mathcal{H})$.
Hence, from the definition Eq.~\eqref{eq:def_stochastic_map} and the monotonicity Eqs.~\eqref{eq:quotient_ineq} and~\eqref{eq:monotonicity_distance}, we obtain
\begin{align}
    &d\left(\xsum_{j} p_j \ket{p_j}\bra{p_j}, \xsum_{j} q_j \ket{p_j}\bra{p_j}\right)\geq d\left(\sum_i r_i \ket{r_i}\bra{r_i},\sum_i s_i \ket{r_i}\bra{r_i}\right)\geq \widetilde{d}\left({\mathbf{r}}^\uparrow,{\mathbf{s}}^\uparrow\right)
    =\widetilde{d}(\widetilde{\mathcal{E}}([\rho]), \widetilde{\mathcal{E}}([\sigma])).
\end{align}
Combining this inequality with Eq.~\eqref{eq:induced_commute}, we obtain Eq.~\eqref{eq:monotonicity_induced_distance}.

\subsection{Proof of convexity [Eq.~\eqref{eq:convexity_induced_distance}]\label{sec:proof_convexity}}
Recall that $d\left(\xsum_{j} p_{i,j} \ket{p_{i,j}}\bra{p_{i,j}}, \xsum_{j} q_j \ket{p_{i,j}}\bra{p_{i,j}}\right)$ does not depend on basis from unitary invariance of $d$ [Eq.~\eqref{eq:axiom_unitary}].
Letting $\ket{p_{j}}:=\ket{p_{1,j}}$, from Eq.~\eqref{eq:quotient_ineq} and the assumption Eq.~\eqref{eq:assumption2}, we obtain
\begin{align}
    &\sum_i \lambda_i \widetilde{d}([\rho_i],[\sigma])=\sum_i \lambda_i d\left(\xsum_{j} p_{i,j} \ket{p_{j}}\bra{p_{j}}, \xsum_{j} q_j \ket{p_j}\bra{p_j}\right)\geq  d\left(\sum_i\lambda_i \xsum_{j} p_{i,j} \ket{p_{j}}\bra{p_{j}}, \xsum_{j} q_j \ket{p_j}\bra{p_j}\right)\nonumber\\
    &\geq \widetilde{d}\left( {\left(\sum_i\lambda_i \mathbf{p}_i^\uparrow\right)}^\uparrow, \mathbf{q}^\uparrow\right)=\widetilde{d}\left( \sum_i\lambda_i \mathbf{p}_i^\uparrow, \mathbf{q}^\uparrow\right)=\widetilde{d}\left(\sum_i \lambda_i[\rho_i],[\sigma]\right),
\end{align}
where we use the defintions Eqs.~\eqref{eq:def_equivalent_sum} and~\eqref{eq:def_equivalent_multiplication} in the last equality.

\section{Derivations and property of unitarily residual measure examples}
\subsection{Derivation of Eq.~\eqref{eq:dsim_bures}\label{sec:deriv_dsim_bures}} 
Let $ |X|:=\sqrt{X^\dagger X}$ for operators $X$. Before the derivation, we prove the following lemma.
\begin{lem}
    Let $\beta,\gamma\geq 0$ and $\kappa >0$. Then, 
    \begin{align}
        \max_{U^\dagger U=\mathbb{I}}\mathrm{Tr}\left[\left|(U\rho U^\dagger)^\beta \sigma^\gamma\right|^\kappa\right]=\xsum_i \left(p_i^\beta q_i^\gamma\right)^\kappa.
        \label{eq:lem_result}
    \end{align}
    \label{lemma1}
\end{lem}
\begin{proof}
Let $\mathfrak{s}_i(X)$ be the $i$-th singular value of operator $X$. Let $\mathbf{x}^\downarrow$ be a sorted vector which is obtained by arranging the components of $\mathbf{x}\in\mathbb{R}^n$ in non-ascending order (i.e., $x_1^\downarrow\geq x_2^\downarrow\geq \cdots \geq x_n^\downarrow$).
    For two vectors $\mathbf{x},\mathbf{y}\in\mathbb{R}^n$ whose components are all nonnegative, {\em weak log-majorization}, denoted by $\mathbf{x}\prec_{\substack{w\\\log}} \mathbf{y}$, is defined as 
\begin{align}
        \prod_{i=1}^k x^\downarrow_i \le \prod_{i=1}^k y^\downarrow_i, \ \text{for all} \ 1\le k \le n.
        \label{eq:def_w_log_majorization}
    \end{align}
    For any two vectors $\mathbf{x},\mathbf{y}\in\mathbb{R}^n$, {\em weak majorization}, denoted by $\mathbf{x}\prec_{w} \mathbf{y}$, is defined as 
\begin{align}
        \sum_{i=1}^k x^\downarrow_i \le \sum_{i=1}^k y^\downarrow_i, \ \text{for all} \ 1\le k \le n.
        \label{eq:def_w_majorization}
    \end{align}
      For $\kappa >0$, $\mathbf{x}\prec_{\substack{w\\\log}} \mathbf{y} \; \rightarrow \; \mathbf{x}^\kappa\prec_{\substack{w\\\log}} \mathbf{y}^\kappa$ follows from Eq.~\eqref{eq:def_w_log_majorization}, where $\mathbf{x}^\kappa$ denotes $(x_1^\kappa, x_2^\kappa, \cdots x_n^\kappa)$. Since $\mathbf{x}\prec_{\substack{w\\\log}} \mathbf{y}$ implies $\mathbf{x}\prec_{w} \mathbf{y}$~\cite{marshall11}, it follows that 
\begin{align}
        \mathbf{x}\prec_{\substack{w\\\log}} \mathbf{y}\; \rightarrow \; \mathbf{x}^\kappa\prec_{w} \mathbf{y}^\kappa.
        \label{eq:log_weak_majorization}
    \end{align}
From the result in Ref.~\cite{marshall11}, it follows that
    \begin{align}
        \prod_{i=1}^k \mathfrak{s}_i^\downarrow(XY) \le \prod_{i=1}^k \mathfrak{s}_i^\downarrow(X)\mathfrak{s}_i^\downarrow(Y)\ \text{for all} \ 1\le k \le n.
        \label{eq:singular_log_majorized}
    \end{align}
    Combining this relation with Eqs.~\eqref{eq:def_w_log_majorization}, ~\eqref{eq:def_w_majorization} and~\eqref{eq:log_weak_majorization}, we obtain
\begin{align}
        \mathrm{Tr}[|XY|^\kappa]=\sum_{i=1}^n \mathfrak{s}_i(XY)^\kappa\le  \xsum_{i=1}^n{\left(\mathfrak{s}_i(X)\mathfrak{s}_i(Y)\right)}^\kappa.
        \label{eq:eq:singular_majorized_phi}
    \end{align}
    Substituting $X=(U\rho U^\dagger)^\beta=U\rho^\beta U^\dagger$
and $Y=\sigma^\gamma$, and combining $\mathfrak{s}_i(X)=p_i^\beta$ and $\mathfrak{s}_i(Y)=q_i^\gamma$ with Eq.~\eqref{eq:eq:singular_majorized_phi}, we obtain 
    \begin{align}
        \mathrm{Tr}\left[\left|(U\rho U^\dagger)^\beta \sigma^\gamma\right|^\kappa\right]\le \xsum_i \left(p_i^\beta q_i^\gamma\right)^\kappa,
        \label{eq:lemma_eq}
    \end{align}
    where we use $({p_i^\beta})^\uparrow=({p_i^\uparrow})^\beta$ and $({q_i^\gamma})^\uparrow=({q_i^\uparrow})^\gamma$.
     Letting $W:= \xsum_{i}\ket{q_i}\bra{p_i}$, it follows that $W$ is a unitary operator such that $W\rho W^\dagger=\xsum_{i} p_i\ket{q_i}\bra{q_i}$. Therefore, we obtain 
     \begin{align}
        \mathrm{Tr}\left[\left|(W\rho W^\dagger)^\beta \sigma^\gamma\right|^\kappa\right]= \xsum_i \left(p_i^\beta q_i^\gamma\right)^\kappa.
        \label{eq:renyi_equality}
    \end{align}
    By combining this equality with Eq.~\eqref{eq:lemma_eq}, we obtain Eq.~\eqref{eq:lem_result}.
\end{proof}
From Eqs.~\eqref{eq:L_D_def} and~\eqref{eq:fidelity_def}, the Bures angle can be written as $\mathrm{arccos}\left(\mathrm{Tr}[|\sqrt{\rho}\sqrt{\sigma}|]\right)$.
By applying Lemma~\ref{lemma1} for $\beta=\gamma=1/2$ and $\kappa=1$, we obtain Eq.~\eqref{eq:dsim_bures}

\subsection{Derivation of Eq.~\eqref{eq:ineq_eigen_trace}\label{sec:deriv_trace_eigen}} 
From the Mirsky inequality, 
\begin{align}
    \|A-B\|_1=\xsum_{i} \mathfrak{s}_i(A-B)\geq \xsum_{i} |\mathfrak{s}_i(A)-\mathfrak{s}_i(B)|
    \label{eq:trace_eigen_ineq}
\end{align}
holds for arbitrary hermitian operator $A$ and $B$~\cite{mirsky1960symmetric,bhatia1996matrix}.
Since singular values of $A$ and $UA U^\dagger$ is equal for a unitary operator $U$, from Eqs.~\eqref{eq:def_metric_quotient} and~\eqref{eq:trace_eigen_ineq}, we obtain 
\begin{align}
    \widetilde{\mathcal{T}}([\rho], [\sigma])\geq \frac{1}{2}\xsum_{i} |p_i - q_i|.
    \label{eq:ineq_forward}
\end{align} 
From $\mathcal{T}(W\rho W^\dagger, \sigma)=1/2\xsum_{i} |p_i-q_i|$, we obtain Eq.~\eqref{eq:ineq_eigen_trace}.

\subsection{Derivation of Eq.~\eqref{eq:induced_renyi_relative_entropy}\label{sec:deriv_induced_renyi_relative_entropy}}
Let $g$ and $h$ be monotonically increasing functions.
Letting $\rho=\sum_i p_i\ket{p_i}\bra{p_i}$ and $\sigma=\sum_j q_j\ket{q_j}\bra{q_j}$, we obtain
\begin{align}
    \mathrm{Tr}[g(U\rho U^\dagger)h(\sigma)]=\sum_{i,j} g(p_i)h(q_j)|\braket{q_j|U|p_i}|^2=\sum_{i,j}C_{ij}g(p_i)h(q_j)=:F(C),
\end{align}
where $C_{ij}:=|\braket{q_j|U|p_i}|^2$.
The matrix $C_{ij}$ is the doubly stochastic matrix (i.e., $\sum_i C_{ij}=\sum_j C_{ij}=1$ and $C_{ij}\geq 0$ for all $i$ and $j$). Since the function $F(C)$ and the constraints of $C_{ij}$ are all linear, the objective function $F(C)$ is maximized when $C_{ij}=1$ or $C_{ij}=0$. By combining these conditions with constraints $\sum_i C_{ij}=\sum_j C_{ij}=1$, we obtain $C_{i\pi(i)}=1$ for all $i$. Here $\pi$ is a permutation of subscripts $\{i\}$. Therefore, we obtain
\begin{align}
    \mathrm{Tr}[g(U\rho U^\dagger)h(\sigma)]\le \sum_i g(p_i)h(q_{\pi(i)})\le \xsum_{i} g(p_i)h(q_i),
    \label{eq:trace_fg_inequality}
\end{align}
where we use 
\begin{align}
    \sum_i a_i b_i \le \xsum_i a_i b_i,
    \label{eq:product_increasing}
\end{align}
for real numbers $\{a_i\}$ and $\{b_i\}$.  
When $\alpha>1$, setting $g(x)=x^\alpha$ and $h(x)=-x^{1-\alpha}$ in Eq.~\eqref{eq:trace_fg_inequality} and using Eqs.~\eqref{eq:def_metric_quotient} and~\eqref{eq:Quantum_renyi_def}, we obtain 
\begin{align}
    \widetilde{D}_\alpha([\rho]\,\|\, [\sigma])\geq \frac{1}{\alpha-1}\ln\left(\xsum_{i} p_i^\alpha q_i^{1-\alpha}\right).
    \label{eq:induced_relative_renyi_inequality}
\end{align}
When $0<\alpha<1$, setting $g(x)=x^\alpha$ and $h(x)=x^{1-\alpha}$ also yields Eq.~\eqref{eq:induced_relative_renyi_inequality}.
From $ D_\alpha(W\rho W^\dagger\| \sigma)=(\alpha-1)^{-1}\ln\left(\xsum_{i} p_i^\alpha q_i^{1-\alpha}\right)$, we obtain Eq.~\eqref{eq:induced_relative_renyi_inequality}.

\subsection{Derivation of Eq.~\eqref{eq:induced_sandwiched}\label{sec:deriv_induced_sandwiched}}
Before the derivation, we prove the following lemma.
\begin{lem}
    Let $\beta\geq 0$, $\gamma< 0$ and $\kappa >0$. Then, 
    \begin{align}
        \min_{U^\dagger U=\mathbb{I}}\mathrm{Tr}\left[\left|(U\rho U^\dagger)^\beta \sigma^\gamma\right|^\kappa\right]=\xsum_i \left(p_i^\beta q_i^\gamma\right)^\kappa.
        \label{eq:lem_result2}
    \end{align}
    \label{lemma2}
\end{lem}
\begin{proof}
    We discuss the differences from Lemma~\ref{lemma1}. 
    From the result in Ref.~\cite{Gelfand:1950:unitary}, it follows that
    \begin{align}
        \prod_{i=1}^k \mathfrak{s}_i^\downarrow(XY) \geq \prod_{i=1}^k \mathfrak{s}_i^\downarrow(X)\mathfrak{s}_i^\uparrow(Y)\ \text{for all} \ 1\le k \le n.
        \label{eq:singular_log_majorized2}
    \end{align}
Note that the statement equivalent to Eq.~\eqref{eq:singular_log_majorized2} can 
be found in Ref.~\cite{marshall11} (H.1.c. Theorem in page 340). 
   Similarly to the derivation of Eq.~\eqref{eq:eq:singular_majorized_phi} , we obtain
\begin{align}
        \mathrm{Tr}[|XY|^\kappa]=\sum_{i=1}^n \mathfrak{s}_i(XY)^\kappa\geq \sum_{i=1}^n {\left(\mathfrak{s}_i^\downarrow(X)\mathfrak{s}^\uparrow_i(Y)\right)}^\kappa.
        \label{eq:eq:singular_majorized_phi2}
    \end{align}
    Substituting $X=(U\rho U^\dagger)^\beta=U\rho^\beta U^\dagger$ and $Y=\sigma^\gamma$, and combining $\mathfrak{s}_i(X)=p_i^\beta$ and $\mathfrak{s}_i(Y)=q_i^\gamma$ with Eq.~\eqref{eq:eq:singular_majorized_phi2}, we obtain 
    \begin{align}
        \mathrm{Tr}\left[\left|(U\rho U^\dagger)^\beta \sigma^\gamma\right|^\kappa\right]\geq \xsum_i \left(p_i^\beta q_i^\gamma\right)^\kappa,
        \label{eq:lemma_eq2}
    \end{align}
    where we use $({q_i^\gamma})^\uparrow=({q_i^\downarrow})^\gamma$.
    From Eq.~\eqref{eq:renyi_equality}, the result follows.
\end{proof}
From Eq.~\eqref{eq:sandwiched_renyi_def}, the sandwiched R\'{e}nyi relative entropy can be written as 
\begin{align}
    D^\prime_\alpha(\rho||\sigma)=\frac{1}{\alpha-1}\ln\left(\mathrm{Tr}\left[\left|\rho^{\frac{1}{2}} \sigma^{\frac{1-\alpha}{2\alpha}}\right|^{2\alpha}\right]\right).
\end{align}
For $\alpha\in (0,1)$, by applying Lemma~\ref{lemma1} for $\beta=1/2$, $\gamma=(1-\alpha)/(2\alpha)\geq 0$ and $\kappa=2\alpha >0$, we obtain Eq.~\eqref{eq:induced_sandwiched}.
Similarly, for $\alpha\in (1,\infty)$, we apply Lemmq~\ref{lemma2} for $\beta=1/2$, $\gamma=(1-\alpha)/(2\alpha)< 0$ and $\kappa=2\alpha>0$.

\section{Derivation of relations in non-hermitian dynamics \label{seq:deriv_nonhermitian}}
\subsection{Derivation of Eq.~\eqref{eq:main_result_MT_like_non_Hermitian} \label{subseq:deriv_nonhermitian_bures}}
We can purify $\widehat{\rho}(0)$ as follows:
 \begin{align}
    \ket{\widehat{\rho}(0)}=\sum_{i}\sqrt{p_{i}(0)}\ket{p_{i}(0)}\otimes\ket{a_{i}},
    \label{eq:def_rho_purify}
\end{align}
where $\{\ket{a_{i}}\}$ are orthonormal basis in the ancilla. 
The time evolution of the purified state in Eq.~\eqref{eq:def_rho_purify} is given by
\begin{align}
    d_t\ket{\widehat{\rho}(t)}=\left\{(-i\mathcal{H}(t)+\braket{\Gamma}(t))\otimes\mathbb{I}_{A}\right\}\ket{\widehat{\rho}(t)},
    \label{eq:purified_Schrodinger_eq}
\end{align}
where $\mathbb{I}_A$ denotes an identity operator of the ancilla.
From Eq.~\eqref{eq:purified_Schrodinger_eq}, letting $X_I(t):= V(t)^\dagger X(t) V(t)$ and $\delta X(t):= X(t)-\braket{X}(t)$, the time evolution of the purified vector in interaction picture 
$\ket{\widehat{\rho}_I(t)}:=V(t)^\dagger\ket{\widehat{\rho}(t)}$ is governed by 
\begin{align}
    d_t\ket{\widehat{\rho}_I(t)}=-\left(\delta\Gamma_I(t)\otimes \mathbb{I}_A\right)\ket{\widehat{\rho}_I(t)}.
    \label{eq:state_vec_non_hermitian}
\end{align}
Recall that the fidelity between purified density matrices can be written as $\mathrm{Fid}(\ket{\rho}, \ket{\sigma})=|\braket{\rho|\sigma}|^2$. Since $\ket{\widehat{\rho}_I(t)}$ is normalized, the Bures angle between $\ket{\widehat{\rho}_I(t)}$ and $\ket{\widehat{\rho}_I(t+dt)}$ can be expanded by
\begin{align}
    \mathcal{L}_D(\ket{\widehat{\rho}_I(t)}, \ket{\widehat{\rho}_I(t+dt)})=\mathrm{arccos}(|\braket{\widehat{\rho}_I(t+dt)|\widehat{\rho}_I(t)}|)=\sqrt{g_{\mathrm{FS}}(t)}dt + O(dt^2),
    \label{eq:bures_fubini}
\end{align}
where $g_{\mathrm{FS}}(t)$ denotes the Fubini-Study metric defined by 
\begin{align}
    g_{\mathrm{FS}}(t):= \braket{d_t \widehat{\rho}_I(t)|d_t \widehat{\rho}_I(t)}-|\braket{d_t \widehat{\rho}_I(t)|\widehat{\rho}_I(t)}|^2,
\end{align}
 where $\ket{d_t \widehat{\rho}_I(t)}$ denotes $d_t \ket{\widehat{\rho}_I(t)}$. 
Substituting Eq.~\eqref{eq:state_vec_non_hermitian} into Eq.~\eqref{eq:bures_fubini} and using the triangle inequality, we obtain
\begin{align}
    \mathcal{L}_D(\ket{{\widehat{\rho}}_I(0)}, \ket{{\widehat{\rho}}_I(\tau)})\le \int_0^\tau \sqrt{g_{\mathrm{FS}}(t)}dt =\int_0^\tau \sqrt{\braket{{\widehat{\rho}}_I(t)|\delta{\Gamma}_I(t)^2|{\widehat{\rho}}_I(t)}}dt = \int_0^\tau \dblbrace{\Gamma}(t)dt,
    \label{eq:MTbound_deriv}
\end{align}
where we use $\braket{{\widehat{\rho}}_I(t)|\delta{\Gamma}_I(t)|{\widehat{\rho}}_I(t)}=0$.
By using $\ket{\widehat{\rho}_I(t)}=V(t)^\dagger\ket{\widehat{\rho}(t)}$ and the monotonicity of the fidelity, we obtain Eq.~\eqref{eq:main_result_MT_like_non_Hermitian}.
\subsection{Derivation of  Eq.~\eqref{eq:eigen_non_Hermitian_short_time}\label{subseq:proof_nonhermitian_tightness}}
We may assume that the magnitude relationship of eigenvalues does not change for infinitesimally small $\delta t>0$. 
From $\sqrt{1+\alpha(t)\delta t + \beta(t)({\delta t})^2 +O(({\delta t})^3)} \sim 1 + \alpha(t)\delta t/2+\beta(t)({\delta t})^2/2 -\alpha(t)^2 ({\delta t})^2/8+O(({\delta t})^3)$, we obtain  
\begin{align}
    &\xsum_i \sqrt{p_i(t)p_i(t+\delta t)}=\sum_i \sqrt{p_i(t)p_i(t+\delta t)}=\sum_i p_i(t)\sqrt{1+\frac{1}{p_i(t)}d_tp_i(t) \delta t + \frac{1}{2p_i(t)}d_t^2p_i(t)({\delta t})^2+O(({\delta t})^3)}\nonumber\\
    &=1-\sum_i \frac{1}{8p_i(t)}\left({d_tp_i(t)}\right)^2({\delta t})^2+O(({\delta t})^3)=1-\frac{1}{8}I(t)({\delta t})^2+O(({\delta t})^3),
\end{align}
where we use $d_t\sum_i p_i=d^2_t\sum_i p_i=0$ in the third equality.
The expansion $\arccos(1-x)=\sqrt{2x}+O(x^{3/2})$ yields
\begin{align}
    &\arccos\left(\xsum_i \sqrt{p_i(t)p_i(t+\delta t)}\right)=\frac{\delta t}{2} \sqrt{I(t)}+O(({\delta t})^2).
\end{align}
Combining this relation with Eqs.~\eqref{eq:dsim_bures} and ~\eqref{eq:main_result_eigen_non_Hemirian}, we obtain Eq.~\eqref{eq:eigen_non_Hermitian_short_time}.
Taking the time derivative of
$p_i(t)=\braket{p_i(t)|\widehat{\rho}(t)|p_i(t)}$ 
and using Eq.~\eqref{eq:nonHermitian_Heisenberg_eq_normalized}, we obtain
\begin{align}
    d_tp_i(t)= -2\braket{p_i(t)|\delta \Gamma(t)|p_i(t)}p_i(t),
    \label{eq:master_eigen_non_hermitian}
\end{align}
where we use $\braket{p_i(t)|d_t p_i(t)}+\braket{d_tp_i(t)|p_i(t)}=0$ from $\braket{p_i(t)|p_i(t)}=1$.
When $\{\ket{p_i(t)}\}$ are eigenvectors of $\Gamma(t)$ for all $i$, we obtain $I(t)=4\dblbrace{\Gamma}(t)^2$ from Eq.~\eqref{eq:master_eigen_non_hermitian}. Hence, the equality holds.

\subsection{Derivation of Eq.~\eqref{eq:purity_non_hermitian} \label{subseq:deriv_nonhermitian_purity}}
Since $|p+q-1| \le |p-1/2|+|q-1/2| \le 1$ for $p,q\in[0,1]$, it follows that  
\begin{align}
     &\xsum_{i} |p_i - q_i|\geq \xsum_{i} \left|(p_i+q_i-1)(p_i - q_i)\right|\geq \left| \xsum_{i}(p_i+q_i-1)(p_i - q_i)\right|\geq \left|\sum_{i} p_i^2 - \sum_{i} q_i^2\right|.
     \label{eq:absolute_square}
\end{align}
From the Cauchy-Schwarz inequality, we obtain
\begin{align}
    &\xsum_{i}|p_i-q_i|= \xsum_{i}|\sqrt{p_i}-\sqrt{q_i}||\sqrt{p_i}+\sqrt{q_i}|\le  \sqrt{\xsum_{i} (\sqrt{p_i}-\sqrt{q_i})^2\xsum_{i} (\sqrt{p_i}+ \sqrt{q_i})^2}\nonumber\\
    &= 2\sqrt{1-{\left(\xsum_{i}\sqrt{p_iq_i}\right)}^2}=2\sin\left(\widetilde{\mathcal{L}}_D([\rho],[\sigma])\right).
\end{align} 
Combining this relation with Eqs.~\eqref{eq:absolute_square} and~\eqref{eq:main_result_eigen_non_Hemirian}, we obtain Eq.~\eqref{eq:purity_non_hermitian}.

\section{Expression of quantum dynamical activity $\mathcal{B}(\tau)$\label{sec:QDA_def}}

The quantum dynamical activity $\mathcal{B}(t)$ considered in Section~\ref{sec:non_hermitian} can be expressed as
\begin{align}
\mathcal{B}(\tau) = \mathcal{A}(\tau) + \mathcal{C}(\tau),
\label{eq:mathcalB_def}
\end{align}
where $\mathcal{A}(\tau)$ is the classical dynamical activity [see Eq.~\eqref{eq:classical_DA_def}] and $\mathcal{C}(\tau)$ represents the contribution from coherent dynamics. The coherent term $\mathcal{C}(\tau)$ is given by \cite{Nishiyama:2024:ExactQDAPRE}
\begin{align}
\mathcal{C}(\tau):= 8\int_{0}^{\tau}ds_{1}\int_{0}^{s_{1}}ds_{2}\mathrm{Re}\left[\mathrm{Tr}\left\{ H_{\mathrm{eff}}^{\dagger}\check{H}\left(s_{1}-s_{2}\right)\rho\left(s_{2}\right)\right\} \right]-4\left(\int_{0}^{\tau}ds\mathrm{Tr}\left[H\rho(s)\right]\right)^{2},
\label{eq:mathcalC_def}
\end{align}
where $\check{H}(t) := e^{\mathcal{L}^{\dagger} t} H$ represents the Hamiltonian $H$ in the Heisenberg picture and the adjoint superoperator $\mathcal{L}^{\dagger}$ is defined as
\begin{align}
\mathcal{L}^{\dagger}\mathcal{O}:= i\left[H,\mathcal{O}\right]+\sum_{m=1}^{N_{C}}\mathcal{D}^{\dagger}\left[L_{m}\right]\mathcal{O}.
\label{eq:mathcalL_dag_def}
\end{align}
Here, $\mathcal{D}^{\dagger}$ is the adjoint dissipator that acts on an operator $\mathcal{O}$ according to
\begin{align}
\mathcal{D}^{\dagger}[L] \mathcal{O} := L^{\dagger} \mathcal{O} L - \frac{1}{2}\left\{L^{\dagger} L, \mathcal{O}\right\}.
\label{eq:mathcalD_dag_def}
\end{align}

\end{widetext}

\begin{acknowledgments}

This work was supported by JSPS KAKENHI Grant Number JP23K24915.

\end{acknowledgments}


\begin{thebibliography}{30}%
    \makeatletter
    \providecommand \@ifxundefined [1]{%
     \@ifx{#1\undefined}
    }%
    \providecommand \@ifnum [1]{%
     \ifnum #1\expandafter \@firstoftwo
     \else \expandafter \@secondoftwo
     \fi
    }%
    \providecommand \@ifx [1]{%
     \ifx #1\expandafter \@firstoftwo
     \else \expandafter \@secondoftwo
     \fi
    }%
    \providecommand \natexlab [1]{#1}%
    \providecommand \enquote  [1]{``#1''}%
    \providecommand \bibnamefont  [1]{#1}%
    \providecommand \bibfnamefont [1]{#1}%
    \providecommand \citenamefont [1]{#1}%
    \providecommand \href@noop [0]{\@secondoftwo}%
    \providecommand \href [0]{\begingroup \@sanitize@url \@href}%
    \providecommand \@href[1]{\@@startlink{#1}\@@href}%
    \providecommand \@@href[1]{\endgroup#1\@@endlink}%
    \providecommand \@sanitize@url [0]{\catcode `\\12\catcode `\$12\catcode `\&12\catcode `\#12\catcode `\^12\catcode `\_12\catcode `\%12\relax}%
    \providecommand \@@startlink[1]{}%
    \providecommand \@@endlink[0]{}%
    \providecommand \url  [0]{\begingroup\@sanitize@url \@url }%
    \providecommand \@url [1]{\endgroup\@href {#1}{\urlprefix }}%
    \providecommand \urlprefix  [0]{URL }%
    \providecommand \Eprint [0]{\href }%
    \providecommand \doibase [0]{https://doi.org/}%
    \providecommand \selectlanguage [0]{\@gobble}%
    \providecommand \bibinfo  [0]{\@secondoftwo}%
    \providecommand \bibfield  [0]{\@secondoftwo}%
    \providecommand \translation [1]{[#1]}%
    \providecommand \BibitemOpen [0]{}%
    \providecommand \bibitemStop [0]{}%
    \providecommand \bibitemNoStop [0]{.\EOS\space}%
    \providecommand \EOS [0]{\spacefactor3000\relax}%
    \providecommand \BibitemShut  [1]{\csname bibitem#1\endcsname}%
    \let\auto@bib@innerbib\@empty
    \bibitem [{\citenamefont {Breuer}\ and\ \citenamefont {Petruccione}(2002)}]{Breuer:2002:OpenQuantum}%
      \BibitemOpen
      \bibfield  {author} {\bibinfo {author} {\bibfnamefont {H.-P.}\ \bibnamefont {Breuer}}\ and\ \bibinfo {author} {\bibfnamefont {F.}~\bibnamefont {Petruccione}},\ }\href@noop {} {\emph {\bibinfo {title} {The theory of open quantum systems}}}\ (\bibinfo  {publisher} {Oxford university press},\ \bibinfo {year} {2002})\BibitemShut {NoStop}%
    \bibitem [{\citenamefont {Seifert}(2012)}]{Seifert:2012:FTReview}%
      \BibitemOpen
      \bibfield  {author} {\bibinfo {author} {\bibfnamefont {U.}~\bibnamefont {Seifert}},\ }\bibfield  {title} {\bibinfo {title} {Stochastic thermodynamics, fluctuation theorems and molecular machines},\ }\href {http://stacks.iop.org/0034-4885/75/i=12/a=126001} {\bibfield  {journal} {\bibinfo  {journal} {Rep. Prog. Phys.}\ }\textbf {\bibinfo {volume} {75}},\ \bibinfo {pages} {126001} (\bibinfo {year} {2012})}\BibitemShut {NoStop}%
    \bibitem [{\citenamefont {Van~den Broeck}\ and\ \citenamefont {Esposito}(2015)}]{VandenBroeck:2015:Review}%
      \BibitemOpen
      \bibfield  {author} {\bibinfo {author} {\bibfnamefont {C.}~\bibnamefont {Van~den Broeck}}\ and\ \bibinfo {author} {\bibfnamefont {M.}~\bibnamefont {Esposito}},\ }\bibfield  {title} {\bibinfo {title} {Ensemble and trajectory thermodynamics: A brief introduction},\ }\href {https://doi.org/10.1016/j.physa.2014.04.035} {\bibfield  {journal} {\bibinfo  {journal} {Physica A}\ }\textbf {\bibinfo {volume} {418}},\ \bibinfo {pages} {6} (\bibinfo {year} {2015})}\BibitemShut {NoStop}%
    \bibitem [{\citenamefont {Parrondo}\ \emph {et~al.}(2009)\citenamefont {Parrondo}, \citenamefont {Broeck},\ and\ \citenamefont {Kawai}}]{Parrondo:2009:Entropy}%
      \BibitemOpen
      \bibfield  {author} {\bibinfo {author} {\bibfnamefont {J.~M.~R.}\ \bibnamefont {Parrondo}}, \bibinfo {author} {\bibfnamefont {C.~V.~d.}\ \bibnamefont {Broeck}},\ and\ \bibinfo {author} {\bibfnamefont {R.}~\bibnamefont {Kawai}},\ }\bibfield  {title} {\bibinfo {title} {Entropy production and the arrow of time},\ }\href {https://doi.org/10.1088/1367-2630/11/7/073008} {\bibfield  {journal} {\bibinfo  {journal} {New J. Phys.}\ }\textbf {\bibinfo {volume} {11}},\ \bibinfo {pages} {073008} (\bibinfo {year} {2009})}\BibitemShut {NoStop}%
    \bibitem [{\citenamefont {\'{E}dgar Rold\'{a}n}\ and\ \citenamefont {Parrondo}(2012)}]{Roldan:2012:EntropyProduction}%
      \BibitemOpen
      \bibfield  {author} {\bibinfo {author} {\bibnamefont {\'{E}dgar Rold\'{a}n}}\ and\ \bibinfo {author} {\bibfnamefont {J.~M.~R.}\ \bibnamefont {Parrondo}},\ }\bibfield  {title} {\bibinfo {title} {Entropy production and {Kullback}-{Leibler} divergence between stationary trajectories of discrete systems},\ }\href {https://doi.org/10.1103/PhysRevE.85.031129} {\bibfield  {journal} {\bibinfo  {journal} {Phys. Rev. E}\ }\textbf {\bibinfo {volume} {85}},\ \bibinfo {pages} {031129} (\bibinfo {year} {2012})}\BibitemShut {NoStop}%
    \bibitem [{\citenamefont {Esposito}\ \emph {et~al.}(2010)\citenamefont {Esposito}, \citenamefont {Lindenberg},\ and\ \citenamefont {Van~den Broeck}}]{Esposito:2010:EntProd}%
      \BibitemOpen
      \bibfield  {author} {\bibinfo {author} {\bibfnamefont {M.}~\bibnamefont {Esposito}}, \bibinfo {author} {\bibfnamefont {K.}~\bibnamefont {Lindenberg}},\ and\ \bibinfo {author} {\bibfnamefont {C.}~\bibnamefont {Van~den Broeck}},\ }\bibfield  {title} {\bibinfo {title} {Entropy production as correlation between system and reservoir},\ }\href {https://doi.org/10.1088/1367-2630/12/1/013013} {\bibfield  {journal} {\bibinfo  {journal} {New J. Phys.}\ }\textbf {\bibinfo {volume} {12}},\ \bibinfo {pages} {013013} (\bibinfo {year} {2010})}\BibitemShut {NoStop}%
    \bibitem [{\citenamefont {Reeb}\ and\ \citenamefont {Wolf}(2014)}]{Reed:2014:Landauer}%
      \BibitemOpen
      \bibfield  {author} {\bibinfo {author} {\bibfnamefont {D.}~\bibnamefont {Reeb}}\ and\ \bibinfo {author} {\bibfnamefont {M.~M.}\ \bibnamefont {Wolf}},\ }\bibfield  {title} {\bibinfo {title} {An improved {Landauer} principle with finite-size corrections},\ }\href {https://doi.org/10.1088/1367-2630/16/10/103011} {\bibfield  {journal} {\bibinfo  {journal} {New. J. Phys.}\ }\textbf {\bibinfo {volume} {16}},\ \bibinfo {pages} {103011} (\bibinfo {year} {2014})}\BibitemShut {NoStop}%
    \bibitem [{\citenamefont {Landi}\ and\ \citenamefont {Paternostro}(2021)}]{Landi:2021:EPReview}%
      \BibitemOpen
      \bibfield  {author} {\bibinfo {author} {\bibfnamefont {G.~T.}\ \bibnamefont {Landi}}\ and\ \bibinfo {author} {\bibfnamefont {M.}~\bibnamefont {Paternostro}},\ }\bibfield  {title} {\bibinfo {title} {Irreversible entropy production: From classical to quantum},\ }\href {https://link.aps.org/doi/10.1103/RevModPhys.93.035008} {\bibfield  {journal} {\bibinfo  {journal} {Rev. Mod. Phys.}\ }\textbf {\bibinfo {volume} {93}},\ \bibinfo {pages} {035008} (\bibinfo {year} {2021})}\BibitemShut {NoStop}%
    \bibitem [{\citenamefont {Van~Vu}\ and\ \citenamefont {Hasegawa}(2021{\natexlab{a}})}]{Vu:2021:GeomBound}%
      \BibitemOpen
      \bibfield  {author} {\bibinfo {author} {\bibfnamefont {T.}~\bibnamefont {Van~Vu}}\ and\ \bibinfo {author} {\bibfnamefont {Y.}~\bibnamefont {Hasegawa}},\ }\bibfield  {title} {\bibinfo {title} {Geometrical bounds of the irreversibility in {Markovian} systems},\ }\href {https://doi.org/10.1103/PhysRevLett.126.010601} {\bibfield  {journal} {\bibinfo  {journal} {Phys. Rev. Lett.}\ }\textbf {\bibinfo {volume} {126}},\ \bibinfo {pages} {010601} (\bibinfo {year} {2021}{\natexlab{a}})}\BibitemShut {NoStop}%
    \bibitem [{\citenamefont {Van~Vu}\ and\ \citenamefont {Hasegawa}(2021{\natexlab{b}})}]{PhysRevLett.127.190601}%
      \BibitemOpen
      \bibfield  {author} {\bibinfo {author} {\bibfnamefont {T.}~\bibnamefont {Van~Vu}}\ and\ \bibinfo {author} {\bibfnamefont {Y.}~\bibnamefont {Hasegawa}},\ }\bibfield  {title} {\bibinfo {title} {Lower bound on irreversibility in thermal relaxation of open quantum systems},\ }\href {https://doi.org/10.1103/PhysRevLett.127.190601} {\bibfield  {journal} {\bibinfo  {journal} {Phys. Rev. Lett.}\ }\textbf {\bibinfo {volume} {127}},\ \bibinfo {pages} {190601} (\bibinfo {year} {2021}{\natexlab{b}})}\BibitemShut {NoStop}%
    \bibitem [{\citenamefont {Van~Vu}\ and\ \citenamefont {Saito}(2023)}]{van2023thermodynamic}%
      \BibitemOpen
      \bibfield  {author} {\bibinfo {author} {\bibfnamefont {T.}~\bibnamefont {Van~Vu}}\ and\ \bibinfo {author} {\bibfnamefont {K.}~\bibnamefont {Saito}},\ }\bibfield  {title} {\bibinfo {title} {Thermodynamic unification of optimal transport: Thermodynamic uncertainty relation, minimum dissipation, and thermodynamic speed limits},\ }\href {https://doi.org/10.1103/PhysRevX.13.011013} {\bibfield  {journal} {\bibinfo  {journal} {Physical Review X}\ }\textbf {\bibinfo {volume} {13}},\ \bibinfo {pages} {011013} (\bibinfo {year} {2023})}\BibitemShut {NoStop}%
    \bibitem [{\citenamefont {Nielsen}\ and\ \citenamefont {Chuang}(2011)}]{Nielsen:2011:QuantumInfoBook}%
      \BibitemOpen
      \bibfield  {author} {\bibinfo {author} {\bibfnamefont {M.~A.}\ \bibnamefont {Nielsen}}\ and\ \bibinfo {author} {\bibfnamefont {I.~L.}\ \bibnamefont {Chuang}},\ }\href {https://doi.org/10.1017/CBO9780511976667} {\emph {\bibinfo {title} {Quantum Computation and Quantum Information}}}\ (\bibinfo  {publisher} {Cambridge University Press},\ \bibinfo {address} {New York, NY, USA},\ \bibinfo {year} {2011})\BibitemShut {NoStop}%
    \bibitem [{\citenamefont {Bhattacharyya}(1946)}]{Bhattacharyya:1946:Divergence}%
      \BibitemOpen
      \bibfield  {author} {\bibinfo {author} {\bibfnamefont {A.}~\bibnamefont {Bhattacharyya}},\ }\bibfield  {title} {\bibinfo {title} {On a measure of divergence between two multinomial populations},\ }\href {http://www.jstor.org/stable/25047882} {\bibfield  {journal} {\bibinfo  {journal} {Sankhy\={a}}\ }\textbf {\bibinfo {volume} {7}},\ \bibinfo {pages} {401} (\bibinfo {year} {1946})}\BibitemShut {NoStop}%
    \bibitem [{\citenamefont {Petz}(1986)}]{Petz:1986:QuasiEntropies}%
      \BibitemOpen
      \bibfield  {author} {\bibinfo {author} {\bibfnamefont {D.}~\bibnamefont {Petz}},\ }\bibfield  {title} {\bibinfo {title} {Quasi-entropies for finite quantum systems},\ }\href {https://doi.org/10.1016/0034-4877(86)90067-4} {\bibfield  {journal} {\bibinfo  {journal} {Rep. Math. Phys.}\ }\textbf {\bibinfo {volume} {23}},\ \bibinfo {pages} {57} (\bibinfo {year} {1986})}\BibitemShut {NoStop}%
    \bibitem [{\citenamefont {M{\"u}ller-Lennert}\ \emph {et~al.}(2013)\citenamefont {M{\"u}ller-Lennert}, \citenamefont {Dupuis}, \citenamefont {Szehr}, \citenamefont {Fehr},\ and\ \citenamefont {Tomamichel}}]{muller2013quantum}%
      \BibitemOpen
      \bibfield  {author} {\bibinfo {author} {\bibfnamefont {M.}~\bibnamefont {M{\"u}ller-Lennert}}, \bibinfo {author} {\bibfnamefont {F.}~\bibnamefont {Dupuis}}, \bibinfo {author} {\bibfnamefont {O.}~\bibnamefont {Szehr}}, \bibinfo {author} {\bibfnamefont {S.}~\bibnamefont {Fehr}},\ and\ \bibinfo {author} {\bibfnamefont {M.}~\bibnamefont {Tomamichel}},\ }\bibfield  {title} {\bibinfo {title} {On quantum r{\'e}nyi entropies: A new generalization and some properties},\ }\href {https://doi.org/10.1063/1.4838856} {\bibfield  {journal} {\bibinfo  {journal} {Journal of Mathematical Physics}\ }\textbf {\bibinfo {volume} {54}} (\bibinfo {year} {2013})}\BibitemShut {NoStop}%
    \bibitem [{\citenamefont {Wilde}\ \emph {et~al.}(2014)\citenamefont {Wilde}, \citenamefont {Winter},\ and\ \citenamefont {Yang}}]{wilde2014strong}%
      \BibitemOpen
      \bibfield  {author} {\bibinfo {author} {\bibfnamefont {M.~M.}\ \bibnamefont {Wilde}}, \bibinfo {author} {\bibfnamefont {A.}~\bibnamefont {Winter}},\ and\ \bibinfo {author} {\bibfnamefont {D.}~\bibnamefont {Yang}},\ }\bibfield  {title} {\bibinfo {title} {Strong converse for the classical capacity of entanglement-breaking and hadamard channels via a sandwiched r{\'e}nyi relative entropy},\ }\href {https://doi.org/10.1007/s00220-014-2122-x} {\bibfield  {journal} {\bibinfo  {journal} {Communications in Mathematical Physics}\ }\textbf {\bibinfo {volume} {331}},\ \bibinfo {pages} {593} (\bibinfo {year} {2014})}\BibitemShut {NoStop}%
    \bibitem [{\citenamefont {R{\'e}nyi}(1961)}]{renyi1961measures}%
      \BibitemOpen
      \bibfield  {author} {\bibinfo {author} {\bibfnamefont {A.}~\bibnamefont {R{\'e}nyi}},\ }\bibfield  {title} {\bibinfo {title} {On measures of entropy and information},\ }in\ \href@noop {} {\emph {\bibinfo {booktitle} {Proceedings of the fourth Berkeley symposium on mathematical statistics and probability, volume 1: contributions to the theory of statistics}}},\ Vol.~\bibinfo {volume} {4}\ (\bibinfo {organization} {University of California Press},\ \bibinfo {year} {1961})\ pp.\ \bibinfo {pages} {547--562}\BibitemShut {NoStop}%
    \bibitem [{\citenamefont {Impens}\ \emph {et~al.}(2021)\citenamefont {Impens}, \citenamefont {d'Angelis}, \citenamefont {Pinheiro},\ and\ \citenamefont {Gu{\'e}ry-Odelin}}]{PhysRevA.104.052620}%
      \BibitemOpen
      \bibfield  {author} {\bibinfo {author} {\bibfnamefont {F.}~\bibnamefont {Impens}}, \bibinfo {author} {\bibfnamefont {F.~M.}\ \bibnamefont {d'Angelis}}, \bibinfo {author} {\bibfnamefont {F.}~\bibnamefont {Pinheiro}},\ and\ \bibinfo {author} {\bibfnamefont {D.}~\bibnamefont {Gu{\'e}ry-Odelin}},\ }\bibfield  {title} {\bibinfo {title} {Time scaling and quantum speed limit in non-{Hermitian} {Hamiltonians}},\ }\href {https://doi.org/10.1103/PhysRevA.104.052620} {\bibfield  {journal} {\bibinfo  {journal} {Phys. Rev. A}\ }\textbf {\bibinfo {volume} {104}},\ \bibinfo {pages} {052620} (\bibinfo {year} {2021})}\BibitemShut {NoStop}%
    \bibitem [{\citenamefont {del Campo}\ \emph {et~al.}(2013)\citenamefont {del Campo}, \citenamefont {Egusquiza}, \citenamefont {Plenio},\ and\ \citenamefont {Huelga}}]{DelCampo:2013:OpenQSL}%
      \BibitemOpen
      \bibfield  {author} {\bibinfo {author} {\bibfnamefont {A.}~\bibnamefont {del Campo}}, \bibinfo {author} {\bibfnamefont {I.~L.}\ \bibnamefont {Egusquiza}}, \bibinfo {author} {\bibfnamefont {M.~B.}\ \bibnamefont {Plenio}},\ and\ \bibinfo {author} {\bibfnamefont {S.~F.}\ \bibnamefont {Huelga}},\ }\bibfield  {title} {\bibinfo {title} {Quantum speed limits in open system dynamics},\ }\href {https://doi.org/10.1103/PhysRevLett.110.050403} {\bibfield  {journal} {\bibinfo  {journal} {Phys. Rev. Lett.}\ }\textbf {\bibinfo {volume} {110}},\ \bibinfo {pages} {050403} (\bibinfo {year} {2013})}\BibitemShut {NoStop}%
    \bibitem [{\citenamefont {Uzdin}\ and\ \citenamefont {Kosloff}(2016)}]{Uzdin:2016:SpeedLimits}%
      \BibitemOpen
      \bibfield  {author} {\bibinfo {author} {\bibfnamefont {R.}~\bibnamefont {Uzdin}}\ and\ \bibinfo {author} {\bibfnamefont {R.}~\bibnamefont {Kosloff}},\ }\bibfield  {title} {\bibinfo {title} {Speed limits in {Liouville} space for open quantum systems},\ }\href {https://doi.org/10.1209/0295-5075/115/40003} {\bibfield  {journal} {\bibinfo  {journal} {EPL}\ }\textbf {\bibinfo {volume} {115}},\ \bibinfo {pages} {40003} (\bibinfo {year} {2016})}\BibitemShut {NoStop}%
    \bibitem [{\citenamefont {Garrahan}(2017)}]{Garrahan:2017:TUR}%
      \BibitemOpen
      \bibfield  {author} {\bibinfo {author} {\bibfnamefont {J.~P.}\ \bibnamefont {Garrahan}},\ }\bibfield  {title} {\bibinfo {title} {Simple bounds on fluctuations and uncertainty relations for first-passage times of counting observables},\ }\href {https://doi.org/10.1103/PhysRevE.95.032134} {\bibfield  {journal} {\bibinfo  {journal} {Phys. Rev. E}\ }\textbf {\bibinfo {volume} {95}},\ \bibinfo {pages} {032134} (\bibinfo {year} {2017})}\BibitemShut {NoStop}%
    \bibitem [{\citenamefont {Shiraishi}\ \emph {et~al.}(2018)\citenamefont {Shiraishi}, \citenamefont {Funo},\ and\ \citenamefont {Saito}}]{Shiraishi:2018:SpeedLimit}%
      \BibitemOpen
      \bibfield  {author} {\bibinfo {author} {\bibfnamefont {N.}~\bibnamefont {Shiraishi}}, \bibinfo {author} {\bibfnamefont {K.}~\bibnamefont {Funo}},\ and\ \bibinfo {author} {\bibfnamefont {K.}~\bibnamefont {Saito}},\ }\bibfield  {title} {\bibinfo {title} {Speed limit for classical stochastic processes},\ }\href {https://link.aps.org/doi/10.1103/PhysRevLett.121.070601} {\bibfield  {journal} {\bibinfo  {journal} {Phys. Rev. Lett.}\ }\textbf {\bibinfo {volume} {121}},\ \bibinfo {pages} {070601} (\bibinfo {year} {2018})}\BibitemShut {NoStop}%
    \bibitem [{\citenamefont {{Di Terlizzi}}\ and\ \citenamefont {Baiesi}(2019)}]{Terlizzi:2019:KUR}%
      \BibitemOpen
      \bibfield  {author} {\bibinfo {author} {\bibfnamefont {I.}~\bibnamefont {{Di Terlizzi}}}\ and\ \bibinfo {author} {\bibfnamefont {M.}~\bibnamefont {Baiesi}},\ }\bibfield  {title} {\bibinfo {title} {Kinetic uncertainty relation},\ }\href {https://doi.org/10.1088/1751-8121/aaee34} {\bibfield  {journal} {\bibinfo  {journal} {J. Phys. A: Math. Theor.}\ }\textbf {\bibinfo {volume} {52}},\ \bibinfo {pages} {02LT03} (\bibinfo {year} {2019})}\BibitemShut {NoStop}%
    \bibitem [{\citenamefont {Hasegawa}(2020)}]{Hasegawa:2020:QTURPRL}%
      \BibitemOpen
      \bibfield  {author} {\bibinfo {author} {\bibfnamefont {Y.}~\bibnamefont {Hasegawa}},\ }\bibfield  {title} {\bibinfo {title} {Quantum thermodynamic uncertainty relation for continuous measurement},\ }\href {https://doi.org/10.1103/PhysRevLett.125.050601} {\bibfield  {journal} {\bibinfo  {journal} {Phys. Rev. Lett.}\ }\textbf {\bibinfo {volume} {125}},\ \bibinfo {pages} {050601} (\bibinfo {year} {2020})}\BibitemShut {NoStop}%
    \bibitem [{\citenamefont {Hasegawa}(2023)}]{Hasegawa:2023:BulkBoundaryBoundNC}%
      \BibitemOpen
      \bibfield  {author} {\bibinfo {author} {\bibfnamefont {Y.}~\bibnamefont {Hasegawa}},\ }\bibfield  {title} {\bibinfo {title} {Unifying speed limit, thermodynamic uncertainty relation and {Heisenberg} principle via bulk-boundary correspondence},\ }\href {https://doi.org/10.1038/s41467-023-38074-8} {\bibfield  {journal} {\bibinfo  {journal} {Nat. Commun.}\ }\textbf {\bibinfo {volume} {14}},\ \bibinfo {pages} {2828} (\bibinfo {year} {2023})}\BibitemShut {NoStop}%
    \bibitem [{\citenamefont {Nishiyama}\ and\ \citenamefont {Hasegawa}(2024)}]{Nishiyama:2024:ExactQDAPRE}%
      \BibitemOpen
      \bibfield  {author} {\bibinfo {author} {\bibfnamefont {T.}~\bibnamefont {Nishiyama}}\ and\ \bibinfo {author} {\bibfnamefont {Y.}~\bibnamefont {Hasegawa}},\ }\bibfield  {title} {\bibinfo {title} {Exact solution to quantum dynamical activity},\ }\href {https://doi.org/10.1103/PhysRevE.109.044114} {\bibfield  {journal} {\bibinfo  {journal} {Phys. Rev. E}\ }\textbf {\bibinfo {volume} {109}},\ \bibinfo {pages} {044114} (\bibinfo {year} {2024})}\BibitemShut {NoStop}%
    \bibitem [{\citenamefont {Marshall}\ \emph {et~al.}(2011)\citenamefont {Marshall}, \citenamefont {Olkin},\ and\ \citenamefont {Arnold}}]{marshall11}%
      \BibitemOpen
      \bibfield  {author} {\bibinfo {author} {\bibfnamefont {A.~W.}\ \bibnamefont {Marshall}}, \bibinfo {author} {\bibfnamefont {I.}~\bibnamefont {Olkin}},\ and\ \bibinfo {author} {\bibfnamefont {B.~C.}\ \bibnamefont {Arnold}},\ }\href {https://doi.org/10.1007/978-0-387-68276-1} {\emph {\bibinfo {title} {Inequalities: Theory of Majorization and Its Applications}}},\ \bibinfo {edition} {2nd}\ ed.,\ Vol.\ \bibinfo {volume} {143}\ (\bibinfo  {publisher} {Springer},\ \bibinfo {year} {2011})\BibitemShut {NoStop}%
    \bibitem [{\citenamefont {Mirsky}(1960)}]{mirsky1960symmetric}%
      \BibitemOpen
      \bibfield  {author} {\bibinfo {author} {\bibfnamefont {L.}~\bibnamefont {Mirsky}},\ }\bibfield  {title} {\bibinfo {title} {Symmetric gauge functions and unitarily invariant norms},\ }\href {https://doi.org/10.1093/qmath/11.1.50} {\bibfield  {journal} {\bibinfo  {journal} {Q. J. Math.}\ }\textbf {\bibinfo {volume} {11}},\ \bibinfo {pages} {50} (\bibinfo {year} {1960})}\BibitemShut {NoStop}%
    \bibitem [{\citenamefont {Bhatia}(1996)}]{bhatia1996matrix}%
      \BibitemOpen
      \bibfield  {author} {\bibinfo {author} {\bibfnamefont {R.}~\bibnamefont {Bhatia}},\ }\href@noop {} {\emph {\bibinfo {title} {Matrix Analysis}}}\ (\bibinfo  {publisher} {Springer New York},\ \bibinfo {year} {1996})\BibitemShut {NoStop}%
    \bibitem [{\citenamefont {Gel'fand}\ and\ \citenamefont {Naimark}(1950)}]{Gelfand:1950:unitary}%
      \BibitemOpen
      \bibfield  {author} {\bibinfo {author} {\bibfnamefont {I.~M.}\ \bibnamefont {Gel'fand}}\ and\ \bibinfo {author} {\bibfnamefont {M.~A.}\ \bibnamefont {Naimark}},\ }\bibfield  {title} {\bibinfo {title} {The relation between the unitary representations of the complex unimodular group and its unitary subgroup},\ }\href@noop {} {\bibfield  {journal} {\bibinfo  {journal} {Izv. Akad. Nauk SSSR Ser. Mat.}\ }\textbf {\bibinfo {volume} {14}},\ \bibinfo {pages} {239} (\bibinfo {year} {1950})}\BibitemShut {NoStop}%
    \end{thebibliography}
\end{document}